\documentclass[10pt]{article}
\usepackage{epsfig}
\usepackage{amssymb}
\usepackage{amsbsy}
\usepackage[colorlinks=true, pdfstartview=FitV, linkcolor=blue, citecolor=blue, urlcolor=blue]{hyperref}

\def\wavepsi{\pi}
\def\wavephi{\rho}
\def\K{\mathcal K}
\def\X {\mathbf X}
\def\Y {\mathbf Y}
\def\bvarphi{{\boldsymbol\varphi}}
\def\bpsi{{\boldsymbol \psi}}
\newcommand\DD{{\cal D}}
\def\wt{\widetilde}

\def\wh{\widehat}

\makeatletter
\@addtoreset{equation}{section}
\makeatother
\def\tr{\mathrm {Tr}}
\def\&{&{\hskip -20pt}}

\def\br{\begin{remark}\small}
\def\er{\end{remark}}

\def\bt{\begin{theorem}}
\def\et{\end{theorem}}
\def\bc{\begin{coroll}}
\def\ec{\end{coroll}}
\def\be{\begin{equation}}
\def\ee{\end{equation}}
\def\bd{\begin{definition}}
\def\ed{\end{definition}}
\def\bp{\begin{proposition}}
\def\ep{\end{proposition}}
\def\bl{\begin{lemma}}
\def\el{\end{lemma}}

\def \pa{\partial}
\def\C{{\mathbb C}}

\def\N{{\mathbb N}}
\def\Z{{\mathbb Z}}
\def\e{{\mathbf e}}
\def\ds{\displaystyle}
\def\res{\mathop{\mathrm {res}}\limits_}

\newtheorem{theorem}{Theorem}[section]
\newtheorem{examp}{Example}[section]
\newtheorem{coroll}{Corollary}[section]

\newtheorem{lemma}{Lemma}[section]
\newtheorem{remark}{Remark}[section]

\newtheorem{proposition}{Proposition}[section]
\newtheorem{definition}{Definition}[section]
\def\ri{\right}
\def\d{{\rm d}}
\def\le{\left}
\def\Amat{\mathbb K}
\def\p{{\mathbf p}}
\def\1{{\bf 1}}
\def\bx{\begin{examp}\small}
\def\ex{\end{examp}}
\def\bea{\begin{eqnarray}}
\def\eea{\end{eqnarray}}
\def\L{\mathcal L}

\def\a{{\alpha}}
\def\Infty{\boldsymbol\infty}
\def \z{\mathbf z}
\date{}
\begin{document}
\fontfamily{cmss}
\fontsize{11pt}{15pt}
\selectfont

\baselineskip 16pt plus 1pt minus 1pt
\begin{titlepage}
\begin{flushright}
CRM-3239\\
\end{flushright}
\vspace{0.2cm}
\begin{center}
\begin{Large}\fontfamily{cmss}
\fontsize{17pt}{27pt} \selectfont \textbf{Effective inverse
spectral problem for rational Lax  matrices and applications}
\end{Large}\\
\bigskip
\begin{large} {M.
Bertola}$^{\dagger\ddagger}$\footnote{Work supported in part by the Natural
    Sciences and Engineering Research Council of Canada (NSERC),
    Grant. No. 261229-03 and by the Fonds FCAR du
    Qu\'ebec No. 88353.}\footnote{bertola@crm.umontreal.ca}
    M. Gekhtman$^\sharp$\footnote{Work supported in part by the NSF grant $\#$ 0400484.}\footnote{mgekhtma@nd.edu}
\end{large}
\\
\bigskip
\begin{small}
$^{\dagger}$ {\em Centre de recherches math\'ematiques,
Universit\'e de Montr\'eal,  C.~P.~6128, succ. centre ville, Montr\'eal,
Qu\'ebec, Canada H3C 3J7} \\
\smallskip
$^{\ddagger}$ {\em Department of Mathematics and
Statistics, Concordia University, 1400 Sherbrooke W., Montr\'eal (QC), H4B 1R6} \\
\smallskip
$^\sharp$ {\em Department of mathematics,  University of Notre Dame,  255 Hurley Hall, Notre Dame, IN 46556-4618}
\end{small}
\end{center}
\bigskip
\begin{center}{\bf Abstract}\\
\end{center}

We reconstruct a rational Lax matrix of size $R+1$ from
its spectral curve (the desingularization of the characteristic polynomial) and some additional data.
Using a twisted Cauchy--like kernel (a bi-differential of bi-weight $(1-\nu,\nu)$)
 we provide a residue-formula for the entries of the Lax matrix in
 terms of bases of dual differentials of weights $\nu,1-\nu$
 respectively.  All objects are described in the most explicit terms using Theta functions.
Via  a sequence of ``elementary twists'', we construct sequences of Lax
matrices sharing the same spectral curve and  polar structure and
related by conjugations by rational matrices. 

Particular choices of  elementary twists lead to construction of
sequences of Lax matrices related to finite--band recurrence relations
(i.e. difference operators) sharing  the same shape. Recurrences of
this kind  are satisfied  by several types of orthogonal and
biorthogonal polynomials. 
The relevance of formul\ae\  obtained to the study of the large degree
asymptotics for these polynomials is indicated.

\medskip
\bigskip
\bigskip
\bigskip
\bigskip

\end{titlepage}
\tableofcontents

\section{Introduction and setting}
The aim of this paper is
to present an explicit solution of the inverse spectral problem for
Lax matrices $A(x)$ of size $(R+1)\times (R+1)$ depending rationally
on $x$. The forward problem and beautiful connections with integrable systems were explored in \cite{BBEI, DKN, MoMu, OPRS, HHA1, HHA2, KrichNov, KrichNov0, KrichNov1a, KrichNov1b, KrichNov1c, HI} (to name a few); in particular it was shown in these works the important r\^ole of the theory of Theta functions \cite{fay} in the solution of both forward and inverse problems.

From the literature cited above we know that the forward problem (under suitable genericity assumptions) produces ``spectral data'' consisting of a smooth algebraic curve $\L$ of genus $g$ with two meromorphic
functions $X, Y:\L\to \C$, where $X$ has degree $R+1$; $X(p)=x$ is the
spectral parameter while $y=Y(p)$ is the eigenvalue of the rational
matrix to be reconstructed. 

The scheme of reconstruction requires that we fix two {\bf dual tensor
  weights} $\nu, 1-\nu$; this means that the eigenvectors of $A(x)$
will be realized as bases in the suitable spaces of sections of
$\nu$--differentials (for the left eigenvectors) and
$(1-\nu)$--differentials for the right eigenvectors. The parameter
$\nu$ can be chosen integer or half-integer. This type of
Baker--Akhiezer functions was considered (but in a different context)
in \cite{OrlovGrinevich}
{ and slightly earlier in the series of papers \cite{KrichNov1a, KrichNov1b, KrichNov1c}.
}
In addition we need a divisor $\Gamma$ (whose degree depends on $\nu$)
and an arbitrary meromorphic differential $\eta$; they will determine
the local properties of the dual BA vectors by fixing the zeroes/poles
and the essential singularity structure, the latter  determined by
$\exp 2i\pi \int \eta$. 

The problem is not new and Baker--Akhiezer functions have been around
for a long time (see, e.g. surveys \cite{Dub, Krich} or the monograph \cite{BBEI}); however our aim is to provide explicit
{\bf residue formul\ae}\ for the entries of $A(x)$ and explore the
intimate relation between a suitable bidifferential $\mathfrak K(p,q)$
of weight $(1-\nu,\nu)$ and several objects of the theory.

We then   ``twist'' the reconstruction scheme so as to obtain recurrence relations for the BA vectors. We
construct a sequence $\Gamma_n$ of divisors of the same degree such
that $\Gamma_{n+1}-\Gamma_n$ is an {\bf elementary divisor} of degree
zero consisting of two arbitrarily chosen points (in general).
{These {\em elementary twists} set in a general framework the original idea}
 {behind the construction of {\bf discrete variable BA functions} suggested in the papers \cite{KrichNov0} and later utilized in \cite{KrichNov} to develop the theory of commuting difference operators.
}

The associated sequence of bidifferentials $\mathfrak K_n$ then
defines a pair of sequences $\wavephi_n, \wavepsi_n$ of
$\nu/(1-\nu)$--differentials (called {\bf dual wave functions}) related by a particular form of {\bf Serre  duality}: such duality is realized via a {\bf residue
  pairing} of the form $\res{\Infty^{(+)}} \wavephi_n\wavepsi_m =
\delta_{mn}$, where $\Infty^{(+)}$ is the divisor of positive points
in the elementary twisting divisors. 

If the elementary twisting divisors are chosen amongst the poles of
$X$ we obtain wave-functions solving a finite--term recurrence
relation  of the form 
\be
X\wavepsi_n = \sum_{j=-d_-}^{d_+} \a_{j}(n) \wavepsi_{n+j}\ .
\ee
These recurrence relations fall within the scope of the theory of
``difference operators'' extensively studied \cite{MoMu, KrichNov}. Our
interest has  a different  origin in connection with
the theory of orthogonal and biorthogonal polynomials and their asymptotics for large
degrees; in this perspective the wave-functions represent
a (formal) asymptotic regime for the polynomials, the  sequence of
bidifferentials $\mathfrak K_n$ is intimately related to
``Christoffel--Darboux-like'' kernels  and
``Christoffel--Darboux-formul\ae'' that arise in those contexts. 


We repeat the argument of \cite{BertoMoAsympt} to
illustrate this connection in the simplest case of the ordinary
orthogonal polynomials. 

Orthogonal polynomials $p_n(x)$ with respect to a weight on the real
line, $w_N (x) = {\rm e}^{-N V(x)} \d x$, satisfy a three--term
recurrence relation\footnote{In fact the situation allows a
  generalization to holomorphic weights on contours as explained in
  \cite{BertoMoAsympt}.} \cite{szego} 
\be
xp_n(x) = \gamma_n p_{n+1}(x) + \beta_n p_n(x) + \gamma_{n-1}p_{n-1}(x)\ ,
\ee
which can be written in matrix form as $x\p = Q\p$, with $\p$ the
semiinfinite vector of the orthogonal polynomials and $Q$ the
tri-diagonal (symmetric) matrix with entries given by  the
coefficients of the above recurrence relations. 

In studying the  large degree asymptotics one  typically sends the
large parameter $N$ appearing in the measure to infinity at the same
rate as the degree $n$ of the polynomial \cite{DKMVZ}: this means that
--while $Q$ implicitly changes because of the change in the measure--
we are considering the polynomials and the recurrence relations very
``far down'' along the diagonal. On a heuristic level one argues that 
the tridiagonal {\bf semi}infinite matrix $Q$ can be replaced by a
{\bf doubly}-infinite matrix $\X$ (i.e. indexed by $\Z$ rather than
$\N$) of the same shape and symmetries. Consider now the associated
functions $\pi_n:= p_n{\rm e}^{-\frac N2 V(x)}$: if $V(x)$  (the {\em
  potential} appearing in the measure that defines the OPs) is a
polynomial of degree $d+1$ then this sequence --while still satisfying
the same three-term recurrence relation-- satisfies also a
$2d+1$--term {\em differential} recurrence relation 
\be
\frac 1 N \pa_x \pi_n = c_d(n) \pi_{n+d} +\dots + -c_d (n-d) \pi_{n-d}
\ee
where in matrix form the recurrence is represented by a
(skew-symmetric)  matrix $P$ with $d$ supra- and $d$
sub-diagonals. The scaling $\frac 1 N$ is needed (on heuristic
grounds) to assure the boundedness of the coefficients of the
recurrence relation. By construction, the two matrices $P,Q$ satisfy 
\be
[P,Q]=\frac 1 N \1
\ee
and in the $N\to\infty$ limit they commute: we thus replace them by
two {\bf commuting} doubly--infinite matrices $\X, \Y$ of the same
shape and symmetries. 
At this point,  the first problem is therefore to classify such pairs
of commuting matrices and much of this has been extensively  analyzed
in \cite{KrichNov}; some additional ingredients (Serre duality) can be
found in \cite{BertoMoAsympt} and are put in a general context in the
present manuscript.

In applications stemming from random matrices, the so--called
Christoffel-Darboux kernel has crucial importance since it generates
all correlation functions \cite{Mehta}. 
The C-D kernel is nothing but the orthogonal projection operator (for
the chosen measure) on the subspace of polynomials of degree $N-1$ or
less and is given by 
\be
K_n(x,x') = \sum_{j=0}^{N-1} p_j(x) p_{j}(x')\ .
\ee
Due to the Christoffel--Darboux theorem it can be expressed in terms of only two OP
\be
K_N(x,x') = \gamma_N \frac {p_{N}(x)p_{N-1}(x') - p_{N-1}(x) p_N(x')}{x-x'}
\ee
and this fact is crucial in proving universality results since it
allows to express the asymptotic behavior for large $N$ in terms of a
fixed (i.e. $N$ independent)  number of polynomials (in this case
$2$). 

In the heuristic approach used in \cite{BertoMoAsympt} (and then
justified rigorously using Riemann--Hilbert techniques) the
quasipolynomials $\pi_n$ were replaced by meromorphic sections of a
spinor bundle, namely by half-differentials on the (asymptotic)
spectral curve, in this case hyperelliptic. The function $x$ was then
regarded as a meromorphic function $X(p)$ on this algebraic curve whose 
multiplication of the half-differentials $\pi_n$ can be
expressed in term of the same sequence of half-differentials, thus
producing a recurrence relation.  The
``orthogonality'' was replaced by a residue pairing between the
sequence $\pi_n$ of half differentials and the Serre--dual sequence
$\pi_n^\star$ of half-differentials\ :  $\res{} \pi_n \pi_n^\star
=\delta_{mn}$. Similarly the kernel $K_n(x,x')$ was replaced by
bidifferential of weights $(1/2,1/2)$ that played the r\^ole of
projection operator with respect to the residue pairing. 

Such bidifferential also satisfies a ``Christoffel--Darboux'' theorem
\be
\mathfrak K_N(p,p') = \gamma_N \frac{\pi_N(p) \pi_{N-1}^\star(p') - \pi_{N-1}(p') \pi_N^\star(p)}{X(p)-X(p')}
\ee
which is ultimately an identity for Theta functions; this is fully
generalized presently in Prop. \ref{CDI} and Prop. \ref{commCDI}. 

Our paper does not focus primarily on difference operators, rather we find them as a byproduct of the sequence of
transformations induced on the Lax matrix by the elementary twisting; also, the eigenvectors for the Lax matrix
(i.e. the Baker--Akhiezer vectors) solve certain Riemann--Hilbert problems with quasi-permutation monodromies.
These were studied in \cite{korotkin} for their own sake, while our
approach finds them as a natural byproduct of the inverse-spectral
reconstruction. 
Riemann--Hilbert problems with quasipermutation monodromies are also
related to asymptotics of (multi)orthogonal polynomials; 
indeed after the so-called normalization of the RH problem satisfied
by the polynomials and associated functions, one is lead 
to an approximating asymptotic problem with quasipermutation monodromies.

\par\vskip 5pt

The paper is organized as follows:
in Section \ref{sectNot} we recall the basic tools from the geometry
of Riemann surfaces, in particular the notion
of Theta functions and prime forms, after \cite{fay}. 

In Section \ref{sectInv} we set up the inverse spectral problem for
rational Lax matrices; here the problem is solved using pairs of dual
Baker--Akhiezer vectors with tensor weights $\nu,1-\nu$ where $\nu \in
\frac 1 2 \Z$. A residue formula for the Lax matrix in terms of
spectral projectors is derived. We also derive the ``generalized Toda
lattice'' in terms of elementary twists and express the ladder
matrices and the matrices implementing the change of a line bundle in
terms of suitable residue formul\ae. Finally, we provide explicit
expressions for the relevant twisted Cauchy kernels in terms of Theta
functions and prime forms. 

In Section \ref{sectFinite} we specialize the generalized Toda lattice
so as to obtain genuine finite-terms recurrence relations (difference
operators); in this setting more explicit formul\ae\ for the BA
vectors are derived. The connection to Riemann--Hilbert problems with
quasi--permutation monodromies is pointed out. 

Finally, in Section \ref{sectComm} we consider the case that is
potentially most relevant to the study of biorthogonal polynomials for
the two--matrix model \cite{BertoEynardHarnad} and reveal a notion of
duality that is well known for biorthogonal polynomials but was  not
known in the context of pairs of commuting difference operators. 

We end this introduction pointing out that in the case that $\L$ has genus $0$ all the formul\ae\ can be expressed in terms of rational functions of the uniformizing parameter: this is left as exercise for the interested reader.

\section{Notation and main tools}
\label{sectNot}
\subsection{Theta functions}
For a given smooth genus-$g$ curve $\L$ with a fixed choice of
symplectic homology basis  of $a$ and $b$-cycles,  we denote by 
$\omega_j$ the normalized basis of holomorphic differentials
\be
\oint_{a_j} \omega_\ell  = \delta_{j\ell}\ ,\qquad \oint_{b_j} \omega_{\ell} = \tau_{j\ell} = \tau_{\ell j}\ .
\ee
We will denote by
$\Theta$ the theta function
\be
\Theta (\z):= \sum_{\vec n\in \Z^g} {\rm e}^{i\pi \vec n\cdot \tau \vec n - 2i\pi \z\cdot \vec n}
\ee
For brevity we will often omit any symbolic reference to the Abel map:
namely if $p\in \L$ is a point and it appears as argument of a
Theta-function, it will be understood  that the Abel map (with a
certain basepoint) was applied.\\ 
We denote by $\K$ the vector of Riemann constants (also depending on the choice of the basepoint)
\be
\K_j = -\sum_{\ell=1}^{g} \le[ \oint_{a_\ell} \omega_\ell(p) \int_{p_0}^p \omega_j(q) - \delta_{j\ell} \frac
{\tau_{jj}}2   \ri]
\ee
where in this expression the cycles $a_j$ are realized as loops with
basepoint $p_0$ and the inner integration is done along a path lying
in the canonical dissection of the surface along the chosen
representatives of the basis in the homology of the curve. 

The crucial property of $\K$ is that for a nonspecial divisor $\Gamma$
of degree $g$, $\Gamma  = \sum_{j=1}^g \gamma_j$, the "function" 
\be
f(p) = \Theta(p - \Gamma - \K)
\ee
has zeroes precisely and only at $p=\gamma_j$, $j=1\dots g$.

We will also have to use Theta functions with (complex)
characteristics; for any two complex vectors $\vec \epsilon, \vec
\delta$ the theta function with these (half) characteristics is
defined via 
\bea
\Theta\le[{\vec \epsilon \atop \vec \delta}\ri] (\z) :=
\exp\le(2i\pi\le( \frac {\epsilon \cdot \tau \cdot \epsilon }8 + \frac
1 2 \epsilon \cdot \z + \frac 1 4 \epsilon \cdot \delta\ri) \ri)
\Theta\le(\z + \frac {\vec \delta} 2 + \tau \frac {\vec \epsilon }2
\ri) 
\eea
Here the (half) characteristics of a point  $\zeta\in \C^g$ are defined by
\be
2\zeta = \delta + \tau \epsilon
\ee
where the factor of $2$ is purely conventional so that half integer
characteristics have integer (half)-characteristics. 
This modified Theta function  has the  following periodicity property :  for $\lambda, \mu\in \Z^g$
\bea
\Theta\le[{\vec \epsilon \atop \vec \delta}\ri] (\z+\lambda  +
\tau\mu)  = \exp\le[ i\pi (\vec \epsilon \cdot \lambda - \vec \delta
\cdot \mu) -i\pi \mu\cdot \tau \cdot \mu  - 2i\pi \z\cdot \mu \ri] 
\Theta\le[{\vec \epsilon \atop \vec \delta}\ri] (\z)
\eea
Note also the symmetry
\be
\Theta\le[{\vec \epsilon \atop \vec \delta }\ri](\z) = \Theta\le[{-\vec \epsilon \atop- \vec \delta }\ri](-\z)
\ee
\subsection{Prime form}
The prime form
$E(\zeta,\zeta')$ is defined as follows \cite{fay}
\bd
The prime form $E(\zeta,\zeta')$ is the $(-1/2,-1/2)$ bi-differential on $\L\times\L$
\bea
E(\zeta,\zeta') = \frac {\Theta_\Delta
  (\mathfrak u(\zeta)-\mathfrak u(\zeta'))}
{h_{\Delta} (\zeta) h_{\le[\alpha\atop \beta\ri]}  (\zeta') }
\\
h_{\Delta} (\zeta)^2 := \sum_{k=1}^{g}
\pa_{\mathfrak u_k}\ln\Theta_\Delta\bigg|_{\mathfrak
  u=0} \omega_k(\zeta)\ ,
\eea
where $\omega_k$ are the normalized Abelian holomorphic differentials,
$\mathfrak u$ is the corresponding Abel map and $\Delta  = \le[\alpha\atop
  \beta\ri]$ is a half--integer odd characteristic (the prime form does
not depend on which one).
\ed
The prime form $E(\zeta,\zeta')$ is antisymmetric in its arguments and
it is a section of an appropriate line bundle, i.e. it is multiplicatively multivalued on $\L\times \L$:
\bea
&&
E(\zeta + a_j,\zeta') = E(\zeta,\zeta')\ ,\qquad
 E(\zeta + b_j, \zeta') = E(\zeta,\zeta') \exp{ \le(-\frac {\tau_{jj}}2  - \int_\zeta^{\zeta'} \omega_j\ri)}
\eea
In our notation for the (half)-characteristics, the vectors
$\alpha,\beta$ appearing in the definition of the prime form are
actually integer valued. 
We also note for future reference that the half order differential
$h_\Delta$ is also multivalued according to 
\bea
h_\Delta(p+a_j) = {\rm e}^{i\pi \alpha_j} h_\Delta(p)\\
h_\Delta(p +b_j) ={\rm e}^{-i\pi \beta_j} h_\Delta(p).
\label{basespinor}
\eea
\section{Inverse spectral problem for rational Lax matrices}
\label{sectInv}
The goal of this section is to explore the inverse spectral problem,
namely how to reconstruct a matrix rationally dependent on $x$ from
the knowledge of its spectral curve and some additional data. 
Since the construction is quite symmetric we can also deal with the
dual situation without any extra effort, thus treating the spectral
parameter and the eigenvalues on the same footing. 

We work with the following data
\begin{enumerate}
\item A smooth curve $\L$ of genus $g$.
\item Two meromorphic functions $X,Y$ with polar divisors $\mathfrak X, \mathfrak Y$   of degrees $R+1$ and $S+1$.
\item A (generic) divisors $\Gamma$ of degree $g+R$ (not necessarily positive).
\end{enumerate}

The main tool is the following adaptation of the Cauchy kernel \cite{fay, KrichNov1a, KrichNov1b, KrichNov1c}
\bp
For a generic choice of divisor $\Gamma$ there exists   a  unique kernel $K(p,\xi)$ which is a
function w.r.t. the point $p$ and a differential w.r.t. the point $\xi$ with the divisor properties
\bea
&& (K(p,\xi))_p \geq -\Gamma + \mathfrak X - \xi\label{div1}\\
&& (K(p,\xi))_\xi \geq \Gamma - \mathfrak X - p\label{div2}
\eea
such that $\res{\xi = p} K(p,\xi) = 1$. The subscripts above indicate
in which variable the divisor properties are considered. 
\ep
The proof follows easily from the Riemann--Roch theorem; we will write
explicitly the expression of this kernel (in a generalized setting of
which the current one is a particular case) 
in terms of Theta functions later on.
\br[Linear equivalence]
In fact we could be slightly more general in the formulation of the
above proposition, since what matters there is only the equivalence
class (modulo principal divisors) of $\Gamma -\mathfrak X$. In
particular we could use in (\ref{div1}, \ref{div2}) two different
divisors $\mathcal D,-\wt{\mathcal D}$ (both of degree $-g-1$) as long
as $\mathcal D$ and $\wt {\mathcal D}$ are equivalent. 

In that case, however, $\res{\xi=p}K(p,\xi)= f(p)$ would be a
meromorphic function with divisor $(f) = \wt{\mathcal  D} - \mathcal
D$; there is only one such function (generically) up to scalar
multiplication. Hence the normalization would have to be fixed in some
other {\em ad hoc} way. 
\er
\bx
If the divisor $\mathfrak X$ consist of $R+1$ distinct points
$x_0,\dots x_R$ and $\Gamma$ is a positive divisor,  then the
expression for $K$ is 
\be
K(p,\xi) =  C\det \le[
\begin{array}{c| ccc}
\rho_{p\infty_0} (\xi) &  \rho_{p\infty_0} (\gamma_1)  & \cdots & \rho_{p\infty_0} (\gamma_{g+R}) \\
\vdots & &&\\
\rho_{p\infty_R} (\xi) &  \rho_{p\infty_R} (\gamma_1)  & \cdots & \rho_{p\infty_R} (\gamma_{g+R}) \\
\hline
\omega_1(\xi) & \omega_1(\gamma_1) & \cdots & \omega_1(\gamma_{g+R})\\
\vdots & & & \\
\omega_g(\xi) & \omega_g(\gamma_1) & \cdots & \omega_g(\gamma_{g+R})
\end{array}\ri]
\ee
where $\rho_{pq}(\xi)$ stands for the (unique) normalized Abelian
differential of the third kind with first--order poles at $p,q$ and
residues $\pm 1$; the constant $C$ depends on the divisors $\mathfrak
X, \Gamma$ and is chosen so that the residue at $\xi=p$ is $1$. 
\ex
Consider now the expression $M(p,\xi):= (X(p)-X(\xi)) K(p,\xi)$; its
divisor properties w.r.t. $p, \xi$ follow from the properties of $K$: 
\bea
&& (M(p,\xi))_p \geq -\Gamma\\
&& (M(p,\xi)_\xi \geq \Gamma - 2 \mathfrak X\ ,
\eea
where the pole on the diagonal is now absent because of the multiplication by $X(p)-X(\xi)$.

Again the Riemann--Roch theorem implies that generically
\be
{\bf r}(-\Gamma) = {\bf i}(\Gamma) -g +\deg \Gamma +1 = R+1
\ee
since (generically) ${\bf i} (\Gamma)=0$.

Also ${\bf i} (\Gamma - 2\mathfrak X) = R+1$ (generically); to see
this we note that the space of third-kind differentials with poles not
exceeding $2\mathfrak X$ has dimension $\deg(2 \mathfrak X)-1 + g =
2R+1+g$. Imposing the vanishing at $g+R$ points gives as many linear
constraints, hence reducing the dimension to $R+1$. 

Let $\psi_0(p), \dots, \psi_R(p)$ be any  basis of the vector space of
meromorphic  functions with divisor exceeding $-\Gamma$ and  let
$\varphi_0(p), \dots, \varphi_R(p)$ be any basis of the vector space
of differentials with divisor  exceeding $\Gamma - 2\mathfrak X$. Let
us introduce the notations 
\be
 \bpsi := \le[\begin{array}{c}\psi_0(p)\\ \vdots\\ \psi_R(p)\end{array}\ri]\ ,\qquad
\bvarphi:= \le[\begin{array}{c} \varphi_0(p)\\ \vdots \\ \varphi_R(p)\end{array}\ri]\ .
\ee
These vectors will be called the {\bf pair of dual  Baker--Akhiezer}
vectors and we will show later on how  this term is motivated by the
Serre duality. 
{ Note that the notion of the {\bf dual Baker-Akhiezer function}
was first introduced in \cite{cher} where it was applied to construct algebro-geometric solutions
to Gelfand-Dickii, NLS and sine-Gordon hierarchies.}

Since the dual pair spans their respective spaces it follows that
\be
M(p,\xi) \in \C\{\psi_0,\dots \psi_R\}\otimes \C\{ \varphi_0,\dots, \varphi_R\}
\ee
In other words there is a $(R+1)\times (R+1)$ matrix  $\Amat$ with constant coefficients such that
\be
(X(p)-X(\xi)) K(p,\xi) = \bvarphi^t(\xi) \Amat \bpsi (p)
\ee
which gives immediately
\bp
\label{CDI}
There is a constant matrix $\Amat$ of size $R+1$, depending on the choice of bases $\psi_j, \varphi_j$, such that
\be
K(p,\xi) = \frac { \bvarphi^t(\xi) \Amat  \bpsi (p)}{X(p)-X(\xi)}
\ee
\ep

From this expression we derive the following identity.
\bc
\label{dX}
Independently of the choices of the bases $ \bpsi,  \bvarphi$ we have the identity
\be
dX(p) = \bvarphi^t(p) \Amat\bpsi(p)\ .
\ee
Moreover, if $p,q\in \mathcal L\setminus (\Gamma \cup \mathfrak X)$ are two points such that $X(p)=X(q)$  then
$\bvarphi^t(p)\Amat \bpsi(q) = 0$.
\ec
{\bf Proof}.
 Taking the residue on the diagonal we have
 \be
 1 = \res{\xi = p}  K(p,\xi) = \frac {\bvarphi^t(p) \Amat \bpsi(p)}{dX(p)}\ .
 \ee
 The second statement follows from the fact that $K(p,q)$ is regular
 and hence the numerator in its expression must vanish whenever $X(p)=X(q)$. {\bf
   Q.E.D.}\par \vskip 5pt 
The Lax matrix can now be constructed from these spectral data if
$\bpsi$ and $\bvarphi^t \Amat$ are viewed  as the right/left
eigenvectors 
with eigenvalue $Y(p)$ at the point(s) $p$ above $x=X(p)$. The
explicit residue formula will be given later in Sect. \ref{sectLax} in
a generalized setting. 
\subsection{Using different tensor weights}
\label{otherweights}
In the above scheme we are using a  pair of BA vectors with tensor weights $0$ and $1$ respectively,
that is, functions  and differentials. This is, in fact unnecessary
and in some applications (typically to the asymptotics of ODEs) it may
even be too restrictive. 

In general,
we could widen the scope of the construction so that $\bpsi$ and
$\bvarphi$ can be tensors of weight $\nu$ and $1-\nu$ respectively; 
similar considerations (motivated by applications to 
 quantum field theory) 
 were used in
\cite{KrichNov1a, KrichNov1b, KrichNov1c, OrlovGrinevich}. The tensor weight $\nu$ can be typically  integer or half--integer;
the particularly useful case \cite{BertoMoAsympt} is $\nu=\frac 12$
where both elements of the pair are spinors
(half--differentials). Clearly 
some modifications in the way the Riemann--Roch theorem is applied are
needed (i. p.,  more general Serre--duality arguments). 

If $h_\nu(\DD)$ is the  dimension of the space of $\nu$--differentials with divisor
exceeding $\DD$ then we know that, for $\nu\geq \frac 1 2 $,
$h_\nu(\DD)=0$ if $\DD$ is generic and $\deg \DD\geq \delta_{\nu 1} +
(2\nu-1)(g-1)$. 

Thus, let $\Gamma$ be generic and of degree $\deg(\Gamma) = (2\nu-1)(g-1)+R+1$; then one finds
from the Riemann--Roch theorem that, for $\nu \in \frac 1 2 \Z, \
j\geq \frac 12 $ (the case $\nu =1$ being the one discussed above) 
\be
h_{1-\nu}(-\Gamma) = h_{\nu}(\Gamma) + \deg(\Gamma) - (2\nu-1)(g-1)  = R+1\ .
\ee
Thus  we should use a divisor $\Gamma$ of  degree $(2\nu -1)(g-1)+R+1$
and  choose $\bpsi$ to be a basis in $H_{1-\nu }(-\Gamma)$ (of
dimension $R+1$). 

The dual BA vector  $\bvarphi$ would then span $H_{\nu }(\Gamma - 2\mathfrak X)$ also of the same dimension.

The relevant ``Cauchy'' kernel is then a bidifferential of weights $(1-\nu, \nu)$ with divisor properties
\be
(\mathfrak K (p,q) )_{\le\{{p\atop q} \ri\}}\geq
\le\{
\begin{array}{c}
-\Gamma + \mathfrak X -q\\
\Gamma - \mathfrak X -p
\end{array}
\ri.
\label{divK}
\ee
and normalized by the requirement that the ``$\nu$--residue'' is one,
namely  the expansion along the diagonal $p=q$ in local coordinate
$z(p)=z,\ z(q)=z'$ is 
\be
\mathfrak K (p,q) = \frac {\d z^{1-\nu}{\d z'}^\nu}{z-z'} (1 + \mathcal O(z-z'))\ .
\ee
The leading coefficient of the expansion is invariant under changes of
a local coordinate and hence it is a geometrical quantity. 

{\bf Note} that if $\nu$ is a half--integer then, for completeness,
one should also choose a spinor bundle (i.e. signs with which the
half-integer spinor changes along each handle). 

Completely similar considerations as before show that
\bp
There exists a constant matrix $\mathbb K^{(\nu)}$ such that
\be
\mathfrak K(p,q) = \frac {\bvarphi(q) \mathbb K^{(\nu)} \bpsi(p)}{X(p)-X(q)}
\ee
and hence
\be
\d X(p) = \bvarphi(p) \mathbb K^{(\nu)} \bpsi(p)
\ee
\ep

In the next sections we can now consider this more general case but
remove the explicit reference to the tensor weight $\nu$. 
Thus we will have
\begin{itemize}
\item $\Gamma$ a generic divisor of degree $(2\nu-1)(g-1)+R+1$;
\item $\bpsi = [\psi_0,\dots, \psi_R]$ a vector of basis
  $(1-\nu)$--differentials in $\mathcal H_{1-\nu} (-\Gamma)$ (of
  dimension $R+1$); 
\item $\bvarphi = [\varphi_0,\dots, \varphi_R]$ a vector of basis
  $\nu$--differentials in $\mathcal H_{\nu}(\Gamma - 2\mathfrak X)$; 
\item $\mathfrak K(p,q)$ the unique bi--tensor of bi--weight
  $(1-\nu,\nu)$ with the divisor properties and the normalization
  listed above. 
\end{itemize}

\subsubsection{Twisting by flat line bundles}
\label{preflat}
In applications to isospectral dynamics and in several other
applications it is necessary to consider a slight generalization of
the above picture, in that 
one twists the line bundle implicitly associated to the divisor
$\Gamma$ by some other line bundle (which may depend on external
parameters or ``times''). 
Typically (as in the classical example of finite-gap integration of KP or KdV dynamics) the extra line-bundle data
falls within the class to be described below.

Let $\eta$ be  an arbitrary meromorphic differential on the curve $\L$ with pole divisor
\be
(\eta) \geq \sum_{i=1}^K d_i c_i, \ ,\ \ c_1,\cdots, c_K  \in \L\ ,\ d_1,\dots, d_K \in \N.
\ee
Let us denote its residues by $t_i$
\be
t_i:= \res{c_i} \eta \ ,\ \ 
\sum_{i=1}^K t_i = 0\ \label{zerores}.
\ee
The Abelian integral $2i\pi \int^p \eta$ (where the base-point of
integration affects only an overall normalization) has in general
nontrivial periods around the $2g$ handles of the curve and around the
punctures $c_i$. The exponential ${\rm e}^{2i\pi\int^p \eta}$ defines
a homomorphism of $\pi_1(\L\setminus\{c_1,\dots,
c_K\})\mapsto\C^\times$ and hence a certain flat line-bundle. 
Moreover this line-bundle has transition functions of exponential
type\footnote{If $d_i=1$ then the singularity may be a pole or power-like singularity with non-integer exponent and hence also branching singularity.} at the punctures $c_i$. 

Twisting the previous description by this line-bundle $\mathfrak
L_\eta$ is then equivalent to considering $\nu$--differentials
(resp. $(1-\nu)$--differentials) with essential singularities at the
punctures $c_i,\ i=1,\dots, K$ of the same type as ${\rm e}^{\pm 2 i
  \pi\int^p \eta}$. 

The {\em twisted} $\nu$--Cauchy kernel is then  a
$(1-\nu,\nu)$--bidifferential with singularities of the form ${\rm
  e}^{2i\pi \int^p\eta}$ and ${\rm e}^{2i\pi\int_q\eta}$, such that
(we still use the same symbol) 
\be
\exp\le(2i\pi \int_q^p\eta\ri) {\mathfrak K}(p,q)
\ee
is locally a $(\nu,1-\nu)$ bidifferential with divisor $\geq
-\Gamma+\mathfrak X-q$ (in $p$) and $\geq \Gamma -\mathfrak X-p$ (in
$q$), and with multiplicative multivaluedness along the homotopy group
of $\L\setminus\{c_1,\dots c_K\}$ given by the character 
\bea
\chi_\eta:\pi_1(\L\setminus\{c_1,\dots,c_K\}) \to \C^\times\cr
\chi_\eta(\gamma) = \exp \le(2i\pi \oint_\gamma \eta\ri)\ .
\eea
 and such that near $c_j$ in a local coordinate $z$ it has a singularity of type $z^{\pm 2i\pi t_j}$.
Thus in general, unless the residues $t_i$ are integers, this kernel has logarithmic branching at the points $c_i$.

Correspondingly, the bases $\bpsi \ (\bvarphi)$ are $1-\nu$--differentials
(resp. $(\nu)$-differentials) with essential singularities of type ${\rm e}^{\pm 2i\pi \int \eta}$. 
The {\bf uniqueness} of such kernel is a simple argument in function
theory and Riemann--Roch theorem. On the existence we do not insist at
this point (although it would not be difficult to prove it abstractly)
since we are going to produce explicit expressions in terms of Theta
functions in Sect. \ref{ThetaCauchy}. 

\subsection{Residue formul\ae\ for the Lax matrix}
\label{sectLax}
Let $\mathfrak J$ be the polar divisor of $Y$. We start with the
observation that $Y(\xi)  \bpsi(\xi) \mathfrak K(p,\xi)$ is a
$1$--differential (in $\xi$) with poles only at $\mathfrak J,
\mathfrak X$ and a simple pole at $p$ with residue $-Y(p) \bpsi(p)$, 
therefore
\bea
Y(p)  \bpsi(p) = -\res{\xi=p} Y(\xi)  \bpsi(\xi) \mathfrak K(p,\xi) =
\sum_{q \in \mathfrak J,\mathfrak X} \res{\xi=q} Y(\xi)  \bpsi(\xi)
\mathfrak K(p,\xi) =\cr 
= \sum_{q \in \mathfrak J,\mathfrak X} \res{\xi=q} Y(\xi) \bpsi(\xi)
\frac {  \bvarphi^t(\xi) \Amat  \bpsi (p)}{X(p)-X(\xi)} = \le[ \sum_{q
  \in \mathfrak J,\mathfrak X} \res{\xi=q} Y(\xi)\frac {   \bpsi(\xi)
  \bvarphi^t(\xi) \Amat}{X(p)-X(\xi)}\ri]  \bpsi (p) 
\eea
The expression
\be
A(x):=  \sum_{q \in \mathfrak J,\mathfrak X} \res{\xi=q} Y(\xi)\frac {   \bpsi(\xi)  \bvarphi^t(\xi) \Amat}{x-X(\xi)}
\label{Laxmatrix}
\ee
is --{\em a priori}-- a rational expression in $x$; it has poles at
the $X$--projection of the divisor $\mathfrak J$ of poles of $Y$ and
at $x=\infty$ (i.e. has a polynomial part). 

In particular  if $\mathfrak y $ is a pole of $Y$ of order $k$ with
$\infty\neq x_o=X(\mathfrak  y)$  and does not coincide with any
branch-point of $X$ (i.e. $\d X\big|_{X^{-1}(x_o)}\neq 0$)  then $A(x)$
has a pole of order $k$ 
\be
A(x) = \mathcal O( (x-x_o)^{-k})
\ee
If $\mathfrak y$ is a branch-point of $X$ and $\mu\geq 1$ is  minimum
order of branching of $X$ ($\mu=2$ being the case of a simple
branch-points) then 
\be
A(x) = \mathcal O \le((x-x_o)^{-[k/\mu]}\ri)
\ee
If $\mathfrak y$ coincides with one of the poles of $X$ then the Lax
matrix will have polynomial parts of degree $k$ or $[k/(d-1)]$ (if $d$
is the order of the pole). Obviously the degree of $A$ depends on the
maximal degree amongst all poles of $Y$ above $x=\infty$. If $Y$ has
no poles coinciding with any of the poles of $X$ then $A(x)$ will be
necessarily bounded at $x=\infty$. 

By construction we have
\be
Y(p) \bpsi(p) = A(X(p))  \bpsi(p)
\ee
Therefore (as expected) $ \bpsi(p)$ is the right eigenvector of the
matrix $A(X(p))$ with eigenvalue $Y(p)$; the different points $p$
lying above the same values of $X(p)$ give the (generically distinct)
eigenvalue/eigenvector pairs. 

\subsubsection{Left eigenvector}
Consider now $Y(p)\mathfrak K(p,\xi)  \bvarphi(p)^t$: this is a
$1$--differential in  $p$ with poles in $p$ at $\mathfrak J, \mathfrak
X$ and $\xi$. Since the $\nu$--residue at $\xi=p$ of $\mathfrak
K(p,\xi)$ is $1$, it follows from a simple computation in a local
coordinate that the residue at $p=\xi$ of this bidifferential is
$Y(\xi)  \bvarphi^t(\xi)$. 
Therefore
\bea
Y(\xi)  \bvarphi^t(\xi) = \res{p=\xi} Y(p)  \bvarphi^t(p)\mathfrak K(p,\xi) =
-\sum_{q\in \mathfrak J, \mathfrak X} \res{p=q}Y(p) \frac {
  \bvarphi^t(\xi) \Amat  \bpsi (p)}{X(p)-X(\xi)}   \bvarphi^t(p)  =\cr 
=    \bvarphi^t(\xi)
\sum_{q\in \mathfrak J, \mathfrak X} \res{p=q}Y(p) \frac {\Amat  \bpsi
  (p)   \varphi^t(p) }{X(\xi)-X(p)} = \bvarphi^t(\xi) \wt A(X(\xi))\\ 
\wt A(x):= \sum_{q\in \mathfrak J, \mathfrak X} \res{p=q}Y(p) \frac {\Amat  \bpsi (p)   \bvarphi^t(p) }{x-X(p)}
\eea

It is clear from the defining formul\ae that
\be
\wt A(x) \Amat  = - \Amat A(x)
\ee
Therefore the left eigenvector of $A(x)$ is $ \bvarphi^t(\xi)\Amat$ :
\be
Y(\xi)  \bvarphi^t(\xi) \Amat = \bvarphi^t(\xi) \wt A(X(\xi)) \Amat =  \bvarphi^t(\xi)\Amat A(x)\bigg|_{x=X(\xi)}\ .
\ee
From  Corollary \ref{dX} it follows that
\be
 {\boldsymbol \ell}(p):= \frac { \bvarphi^t(p)\Amat}{dX(p)}
\ee
is the {\bf normalized} left eigenvector ( a $\nu-1$-differential) of $A(x)$, in the sense that
\be
{\boldsymbol \ell}(p) \cdot\bpsi(p) \equiv 1.
\ee
From the second part of Corollary \ref{dX} follows also that (as it
should) the evaluation of $\ell(q)$ at  the  other points $q\in \L$
above  $X(p)$ are orthogonal to  $ \bpsi(p)$. 
Note that the dual left-eigenvector has
poles at the branch-points of the $X$ projection.

The (generically) rank--one projector on the eigenspace with eigenvalue $Y(p)$ is given by
\be
\Pi(p) := \bpsi(p) \otimes  {\boldsymbol \ell}(p) = \frac { \bpsi(p)  \bvarphi^t(p) \Amat}{dX(p)}
\ee

\subsubsection{Structure of the Lax matrix near a branch-point of $X$}
Suppose $c$ is a critical value of $X$ and $\sum \mu_i \xi_i = X^{-1}(c)$.
Let us choose local coordinates near the point $\xi_i$ as
\be
z_i= (X-c)^{\frac 1{\mu_i}}\ .
\ee
Then we have
\be
L(c) = \res{\mathfrak X, \mathfrak J} \frac {Y(\xi)\Pi(\xi)}{X(\xi)-c} =
\sum_i \res{\xi_i} \frac {Y(\xi) \Pi(\xi)}{z_i(\xi)^{\mu_i}}
\ee
We see that each residue extracts the $\mu_j$ jet of $Y$ and $\Pi$, contributing
to a rank-$\mu_j$ Jordan block of the Lax matrix and this point.

Indeed, by expanding
\be
Y(\xi)  = \sum Y_{j,\ell} {z_j}^\ell\\
\Pi(\xi) = \sum \Pi_{j,\ell} {z_j}^\ell {\rm d}z_j\\
\res{z_j=0} \frac {Y(\xi)\Pi(\xi)}{{z_j}^{\mu_j}} = \sum_{\ell=0}^{\mu_j-1}
Y_{j,\mu_j-1-\ell} \Pi_{j,\ell}
\ee
If $\Pi(z)dz = R(z) L(z) dz$, one sees easily by induction that the rank of any
linear combination of the first $k$ derivatives of $\Pi$ is less or equal to
$k$, and generically is precisely $k$.
\subsubsection{Change of divisor}
\label{sectspec}
The matrix $A(x)$ depends implicitly on the divisor $\Gamma$ but its
characteristic polynomial does not (since the latter describes the
algebraic relation between the rational functions $X, Y$ on the
spectral curve $\L$). We investigate how a change in the divisor
$\Gamma$ (within the same class of generic divisors) affects the Lax
matrix $A(x)$. 

\bp
\label{transition}
Let $A_\Gamma(x)$ the Lax matrix constructed as in Section
\ref{sectLax} using a divisor $\Gamma$. Let $\wt \Gamma$ be another
divisor of the same degree and with similar genericity properties;
then 
\be
A_{\wt \Gamma} (x) = C_{\Gamma,\wt \Gamma} (x) A_{\wt \Gamma}(x) C_{\Gamma, \wt\Gamma}^{-1}(x)
\ee
where
\bea
C_{\Gamma, \wt\Gamma}(x) &\& = \res{\wt \Gamma, \mathfrak X} \frac {\wt \bpsi(\xi) \bvarphi^t(\xi)\Amat }{x-X(\xi)}\\
C_{\wt \Gamma,\Gamma}(x) &\& = \res{ \Gamma, \mathfrak X} \frac {
  \bpsi(\xi) \wt \bvarphi^t(\xi)\wt \Amat }{x-X(\xi)} = C_{\Gamma,\wt
  \Gamma}(x)^{-1} 
\eea
\ep
{\bf Proof.}
The product $\wt \bpsi(\xi) \mathfrak K(p,\xi)$ has poles in $\xi$
only at $\wt \Gamma$, $\mathfrak X$  and at $\xi=p$ with residue $
-\wt \bpsi(p)$. Hence --as before-- 
\be
\wt \bpsi(p) = \res{\wt \Gamma, \mathfrak X} \wt \bpsi(\xi) \mathfrak
K(p,\xi) = \res{\wt \Gamma, \mathfrak X} \frac {\wt \bpsi(\xi)
  \bvarphi^t(p)\Amat }{X(p)-X(\xi)} \bpsi(p) 
\ee
from which the expression for $C_{\Gamma,\wt \Gamma}(x)$ follows. The
expression for $C_{\wt \Gamma, \Gamma}(x)$ and the fact that it is the
inverse of  $C_{\Gamma,\wt \Gamma}(x)$ follows from simply
interchanging the r\^oles of $ \Gamma$ and $\wt \Gamma$. 
Analogous expressions can be found for the change of divisor for the
matrix $\wt A_\Gamma(x)$ (using the basis of forms $\bvarphi$). 
{\bf Q.E.D.}\par \vskip 5pt

Note that the {\em transition matrices} $C_{\Gamma, \wt \Gamma}(x)$
are rational functions of $x$ with poles only at the $X$--projection
of the divisor $\wt \Gamma$ and with a polynomial part of degree equal
to the degree of the subdivisor of  $\wt \Gamma$ that coincides with
some poles of $X$. 
\subsubsection{Change of line bundle}
The matrix $A(x)$ also depends on the chosen third--kind differential
$\eta$ and we investigate what happens when we vary the differential
within the same class. 

Namely, let  $ \wt\eta$ be another third kind differential and
let $\wt \bvarphi, \wt\bpsi, \wt {\mathbb K}, \wt{\mathfrak K}$ be the
same objects constructed  before but using the differential $\wt\eta$
instead of $\eta$.

The one-form  (in $q$)
\be
\mathbb K(p,q) \wt \bpsi(q)
\ee
has poles at $p=q$ with residue  $-\wt \bpsi(p)$, at $\mathfrak X$ and
essential singularities at the poles of $\eta, \wt \eta$.
The transition matrix is then given  by deformation of contours
\be
\wt \bpsi(p)  = \res{q\in \mathfrak X\cup \{c_j,\wt c_j\}} \mathbb K(p,q) \wt \bpsi(q) =
 \res{q\in \mathfrak X\cup \{c_j, \wt c_j\}}  \frac {\wt \bpsi(q)
   \bvarphi^t(q)\mathbb K}{X(p)-X(q)} \bpsi(p) =
 M_{\eta,\wt\eta}(X(p)) \bpsi(p) 
 \ee
 with
 \be
 M_{\eta,\wt\eta}(x) = \res{q\in \mathfrak X\cup \{c_j, \wt c_j\}}
 \frac {\wt \bpsi(q) \bvarphi^t(q)\mathbb K}{x-X(q)} 
 \ee
Here by the symbol $\res{}$ we simply mean  the integral around  a
small loop encircling the point (the differential is not meromorphic
there but has essential singularities). 
The matrix $M_{\eta,\wt\eta}(x)$ has thus essential singularities (in general) in the complex $x$--plane  at the points $X(c_j), X(\wt c_j)$.

Consequently the Lax matrices are related by a simple conjugation.
\be
A_\eta(x) = M_{\eta,\wt \eta} (x)^{-1} A_{\wt \eta}(x) M_{\eta,\wt \eta}(x)
\ee

Note that --by construction-- the Lax matrices are still rational with
the same pole structure even if related by a conjugation with a
non-rational matrix. 
These formul\ae\ provide the integration of any isospectral dynamics
on rational matrix-valued functions available in the literature on
integrable systems 
(see, e.g.  \cite{DKN, HHA1, HHA2, OPRS} and references therein) and
are, in fact, even more general. Indeed if $\eta$ depends (smoothly)
on one or several  a ``time'' parameters, the above formul\ae\ would
provide integration of the flow on the isospectral manifold induced by
the dependence of $\eta$.

Note that the framework we are proposing is more general than the one
in \cite{HHA1, HHA2} since the differential $\eta$ need not have
poles coinciding with any of the poles of $X$ (which is the case in
loc. cit.). 

%
%
\subsubsection{Linear (smooth) deformations of  the line bundle $\chi_\eta$}
 If $\eta$ depends {\em linearly} (or even smoothly) on a set of times generically denoted by $t$
then $\dot M_{\eta} M_{\eta}^{-1}$ is a {\bf rational} matrix (i.e. without essential singularities) as long as the residues of $\eta_t$ are independent of $t$. 

To show this suppose $\eta = \eta_t$ depends smoothly on a parameter $t$; in the literature the dependence is taken to be linear in the sense that the coefficients of the singular parts at the poles (in some chosen and fixed local coordinate) evolve linearly in $t$ but the statement we are making here is more general in that it may include {\bf any} deformation, including a motion of the position of the poles.

Let $\bpsi_t, \bvarphi_t$ be the dual bases of sections evolving in a smooth way; the reader should realize that this evolution implies a ``gauge arbitrariness''  consisting in the freedom of (smooth) change of basis within the same vector spaces of sections. This arbitrariness makes no difference on the rational nature of the infinitesimal deformation.

Denoting by a dot the ``time'' derivative we note that $\dot \bpsi$ (and $\dot \bvarphi$) are then sections of the same tensor space with additional singularities. Indeed, near any of the poles $c_j$ of $\eta$ we have $\bpsi = f_t(p) {\rm e}^{\int^p \eta_t}$, with $f_t(p)$ analytic near $c_j$
\be
\dot\bpsi = \le(\int^p \dot \eta_t f_t + \dot f_t\ri){\rm e}^{\int^p \eta_t}\ .
\ee
Note that $\int^p \dot \eta_t$ near a pole of $\eta_t$ has a pole singularity (without logarithmic term) because the residues of $\eta_t$ are independent of $t$ and hence $\dot \eta_t$ is a second-kind Abelian differential.

Applying the argument that have already used several times before, we obtain
\be
\dot \bpsi_t (p) = -\res{q=p} \mathfrak K_t(p,q) \dot \bpsi_t(q)\ .
\ee
The expression we are taking the residue of, is a differential with poles only along $\mathfrak X + p$ and along the divisor of poles of $\eta_t$ --which we denote by $\mathfrak C$--, due to  $\dot \eta_t$. 
Thus
\be
\dot \bpsi_t (p) = -\res{q=p} \mathfrak K_t(p,q) \dot \bpsi_t(q)  = \res{q\in \mathfrak X+\mathfrak C}\frac{ \dot \bpsi_t(q) \bvarphi_t(q) \mathbb K_t}{X(p)-X(q)} \bpsi_t(p).
\ee
Thus 
\be
\dot M_t(x) M_t(x)^{-1} = \res{q\in \mathfrak X+\mathfrak C}\frac{ \dot \bpsi_t(q) \bvarphi_t(q) \mathbb K_t}{x-X(q)} 
\ee
Since the differential in the numerator is {\bf meromorphic} (without essential singularities) on the spectral curve $\L$, the latter expression is rational in $x\in \C$.
\subsection{Spectral bidifferential}
Suppose we are given a rational $(R+1)\times(R+1)$ matrix $A(x)$ and
let us denote by $y_a(x)$ its (generically simple) $R+1$ eigenvalues;
consider the following bidifferential 
\be
S((x,y_a(x)); (x',y_b(x'))) := \frac {dx\, dx'}{(x-x')^2} \frac {\tr(
  \wt {A-y_a})(x) (\wt{A-y_b})(x'))} {\tr (( \wt{A-y_a})(x)) \tr ((\wt
  {A-y_b})(x'))}\label{specbidi} 
\ee
where the tilde denotes the matrix of co-factors (the classical adjoint).
Since
\be
\Pi_a(x):= \frac {(\wt {A-y_a})(x) }{\tr (( \wt{A-y_a})(x))}
\ee
is the (generically rank--one) projection onto the eigenspace with
eigenvalue $y_a(x)$, it is not difficult to see \cite{BertoMo}
that this bidifferential extends naturally to the spectral curve and,
in fact, we are now going to prove this in general. 
This object has appeared in the context of
isomonodromic deformations of rational connections in the sense of
\cite{JMU}; indeed in \cite{BertoMo} it was shown that  this
bidifferential is the generating function of the Hessian of the
logarithm of the isomonodromic tau function. 

Here we are not considering such deformations but we can relate easily
this bidifferential to the $\nu$--Cauchy kernel introduced above. 
Let $p,q$ be the abstract points on $\L$ with coordinates $(x,y_a(x))$ and $(x',y_b(x'))$.
Note that the expression (\ref{specbidi}) seems to have  at first
sight a double pole whenever two points project to the same $x$-value,
but in fact this occurs only when the branch of the eigenvalue is the
same, meaning that it is a double pole only on the diagonal of the
symmetric product of the spectral curve with itself. 

Indeed it follows immediately that since $\Pi_a(x)\d x =
\bpsi(p)\bvarphi^t(p) \mathbb K$ (and letting $x=X(p), x'=X(q)$) 
\be
S(p,q) = \tr \le(\frac {\bpsi(p) \bvarphi^t (p)\Amat}{X(p)-X(q)}\frac {\bpsi(q) \bvarphi^t (q)\Amat}{X(p)-X(q)}\ri)
=
 \frac { \bvarphi^t (p)\Amat\bpsi(q)} {X(p)-X(q)}
 \frac {\bvarphi^t (q)\Amat\bpsi(p) }{X(p)-X(q)} = \mathfrak K(p,q)\mathfrak K(q,p)
\ee
so that the spectral bidifferential is nothing but the symmetric square of the $\nu$--Cauchy kernel.
It will be shown that $S(p,q)$ is the square of the Szeg\"o\ kernel in Cor. \ref{sega};  notice that,
for the time being, the symmetric square has only a double pole on the
diagonal $p=q$ and no other singularities (this follows from the
divisor properties of $\mathfrak K(p,q)$, (\ref{divK}) and the type of
essential singularities. 

\subsection{Elementary twisting lattice and dual wave functions}
\label{gentoda}
Suppose that we specify a sequence of {\bf elementary} divisors
$\mathfrak T_n$ of degree $0$, $n\in \Z$; by ``elementary'' we mean
that they are of the form 
\be
\mathfrak T_n = \infty^{(+)}_n  - \infty^{(-)}_n
\ee
where $\infty^{(+)}_n, \infty^{(-)}_n$ are two  sequences of points
arbitrarily (but generically) chosen. We will assume that, for any $m$
and $n$, 
$\infty^{(+)}_m \ne\infty^{(-)}_n$'s are distinct from each other (but
points within the same sequence may be repeated).

 If we twist the ``initial'' divisor $\Gamma$
\bea
\Gamma_0&\& := \Gamma\cr
\Gamma_n &\& :=\le\{
\begin{array}{cc}
 \Gamma_{n-1} + \mathfrak T_n  & n \geq 1\\
 \Gamma_{n+1} - \mathfrak T_{n+1} & n \leq -1
 \end{array}\ri.\label{divisToda}
\eea
we obtain a sequence of divisors $\Gamma_n$ of degree $(2\nu-1)(g-1)+R+1$.
Same strategy as before can still be applied (generically), namely we
have a corresponding sequence of bases $\bpsi_n$ and $\bvarphi_n$ and
of Christoffel--Darboux kernels ${\mathfrak K}_n(p,\xi)$ all
satisfying 
\be
{\mathfrak K}_n(p,q)  =  \frac {\bvarphi_n^t(q) \Amat_n \bpsi_n(p)}{X(p)-X(q)}\ ,\qquad
({\mathfrak K}_n(p,q))_{{p\atop q}}\geq \le\{
\begin{array}{c}
-\Gamma_n + \mathfrak X-q\\
\Gamma_n - \mathfrak X -p
\end{array}
\ri.
\label{Kdiv1}
\ee

Let us define $\Infty^{(+)}$ as the {\bf set} of all points
$\infty^{(+)}_n$ (counted without multiplicity with which the may
appear in our sequence). 
\bd
The divisor $\Infty^{(+)} = \bigcup \{\infty^{(+)}_n\}$ will be called the {\bf dualization divisor}.
\ed

\bp
\label{selfrepr}
The sequence of kernels ${\mathfrak K}_n(p,q)$ satisfies
\be
\res{\xi\in \Infty^{(+)}} {\mathfrak K}_n(p,\xi) {\mathfrak K}_m(\xi,q) = \le\{
\begin{array}{cc}
0 &  m\leq n\\
{\mathfrak K}_{m} (p,q) - {\mathfrak K}_n(p,q) & m> n
\end{array}
\ri.
\ee
\ep
The proof is a  simple inspection of the residues; the product of the
two kernels is a differential in $\xi$ that has no poles on
$\Infty^{(+)}$ if $n\geq m$; and if $n<m$ then it  has only poles at a
finite number of points of $\Infty^{(+)}$ and at $\xi=p,\xi=q$, where
it has the indicated residues.

Consider now the difference ${\mathfrak K}_{n+1}(p,q) - {\mathfrak
  K}_n(p,q)$; since both kernels have $\nu$--residue $1$ on the
diagonal, this difference is regular there. 
Moreover
\bea
({\mathfrak K}_{n+1}(p,q) - {\mathfrak K}_n(p,q))_p &\& \geq -\Gamma_n - \infty^{(+)}_{n+1} + \mathfrak X\\
({\mathfrak K}_{n+1}(p,\xi) - {\mathfrak K}_n(p,q))_q &\& \geq \Gamma_n - \infty^{(-)}_{n+1}  -\mathfrak X 
\eea
The two divisors on the right hand side of these inequalities have
degree $-(2\nu-1)(g-1)-1$ and $(2\nu-1)(g-1)-1$ respectively; it
follows that (generically) there is a unique meromorphic
$(1-\nu)$--differential  $\wavepsi_n$ and a unique meromorphic
$\nu$--differential $\wavephi_n$ in the respective spaces specified by
these divisors. We
have proved that 
\bc
\label{telescope}
The difference of two consecutive Christoffel--Darboux kernels in the
generalized Toda  sequence factors 
\be
{\mathfrak K}_{n+1}(p,q) - {\mathfrak K}_n(p,q)  = \wavepsi_n(p) \wavephi_n(q),
\ee
with $\wavepsi_n(p), \wavephi_n(q)$ defined (up to multiplicative constants) by the requirements 
\bea
(\wavepsi_n(p)) &\& \geq -\Gamma_n - \infty^{(+)}_{n+1} + \mathfrak X\label {pidiv}\\
(\wavephi_n(q)) &\& \geq \Gamma_n - \infty^{(-)}_{n+1}  -\mathfrak X \label{rhodiv}
\eea
By induction,
\be
{\mathfrak K}_{n+L} (p,q) - {\mathfrak K}_n(p,q) = \sum_{j=n}^{n+L-1} \wavepsi_j(p) \wavephi_j(q).
\ee
\ec
\br
The name of ``generalized Toda lattice'' is due to the fact that if $X$ is the projection of a hyperelliptic curve (with two simple poles $\L\ni \infty_\pm$ above $x=\infty$) and we use the sequence of elementary divisors $\mathfrak T_n  = \infty_+-\infty_-$ then we recover a setting of the standard Toda lattice theory by looking at suitable isospectral evolution.
\er
Using the Christoffel--Darboux kernels  ${\mathfrak K}_n$ one can
therefore reconstruct a sequence of Lax matrices $A_n(x)$ all sharing
the same spectral curve and connected by conjugation  by the
transition matrices $C_{n, m}(x) := C_{\Gamma_n, \Gamma_m}(x)$
introduced in Prop. \ref{transition}; these {\bf ladder} matrices
satisfy the  obvious relations (which entail discrete  integrability) 
\bea
&& C_{n,m}(x) C_{m,\ell}(x)  = C_{n,\ell}(x)\ ,\qquad \forall n,m,\ell \in \Z.\\
&& C_{n,m}(X(p)) \bpsi_n(p) = \bpsi_m(p)\\
&& \bvarphi^t_n (p) \Amat_n C_{n,m}(X(p)) = \bvarphi^t_m(p) \Amat_m
\eea
%
\br
The previous construction is a generalization of the ``discrete variable Baker-Akhiezer function'', an idea originally formulated in \cite{KrichNov0} which is the hinge of the theory of commuting difference operators.
\er

\subsubsection{Dualization}
Note that
\be
\res{\Infty^{(+)}} \wavephi_m \wavepsi_n = \delta_{mn}
\ee
since if $m\neq n$ the product is a  differential with a polar divisor
of degree $2$ supported  only at the points $\infty^{(+)}_n$'s or only
at the points $\infty^{(-)}_n$'s; only if $m=n$ one pole is at
$\infty^{(+)}_n$ and one at $\infty^{(-)}_n$ so that the residue over
all $\Infty^{(+)}$ is nonzero. The fact  that this residue is actually
1 
follows from Prop. \ref{selfrepr}
\be
\res{\Infty^{(+)}} \le({\mathfrak K}_{n+1}(p,\xi) - {\mathfrak
  K}_n(p,\xi)\ri)\le({\mathfrak K}_{m+1}(\xi,q) - {\mathfrak K}_{m}
(\xi,q) \ri) = \delta_{mn}  \le({\mathfrak K}_{n+1}(p,q) - {\mathfrak
  K}_n(p,q)\ri) 
\ee
which implies that $\res{\Infty^{(+)}} \wavephi_n\wavepsi_n =1$.

In addition we have
\bea
\res{\Infty^{(+)} } \wavephi_{m}(p) {\mathfrak K}_n(p,q) = \le\{
\begin{array}{cc}
0 & m\geq n\\
\wavephi_{m} & m<n
\end{array}
\ri.\ \ \
\res{\Infty^{(+)} } \wavepsi_{m}(q) {\mathfrak K}_n(p,q) = \le\{
\begin{array}{cc}
-\wavepsi_{m} & m\geq n\\
0& m< n
\end{array}
\ri.\label{projphipsi}
\eea

\subsection{Expressions in terms of Theta functions}
\label{ThetaCauchy}
We now present explicit expressions of all the objects introduced so far.
Let us decompose the divisor $\Gamma$ (of degree $(2\nu-1)(g-1)+R+1$) as
\be
\Gamma = \gamma_1+\dots +\gamma_{R+1}
 + \sum_{\ell=1}^{2\nu-1} \Gamma_0^{(\ell)} \ee
where $\Gamma_0^{(\ell)}$ are divisors of degree $g-1$ and
$\gamma_1,\dots, \gamma_{R+1}$ are $R$ points singled out arbitrarily. 

Let $\xi_1,\dots \xi_{2\nu-1}$ be arbitrary fixed points (the final
formul\ae\ will only have a fictitious dependence on them).

\bt
\label{thm_nu_Cauchy}
The twisted $\nu$--Cauchy kernels of Sect. \ref{otherweights} are given by
\be
\mathfrak K_n(p,q) = \frac {T_n(q)}{T_{n}(p)} \frac {F_{\eta,n}(p,q)}{E(p,q)} {\rm e}^{-2i\pi \int_q^p \eta}
\ee
where
\bea
F_{\eta,n}(p,q)&\&:=
\frac{ \Theta\le[{\mathcal A \atop \mathcal B}\ri] (p-q - \Gamma_n- (2\nu-1) \K + \mathfrak X  )}
{ \Theta\le[{\mathcal A \atop \mathcal B}\ri] (   \mathfrak X - \Gamma_n-(2\nu-1) \K )}
\prod_{j=1}^K
\le(\frac{ \Theta_\Delta(q-c_j)
}{ \Theta_\Delta(p-c_j)}\ri)^{-2i\pi t_j}
\label{Feta}\\
T_n(q) &\& =
\prod_{\ell=1}^{2\nu-1}
\frac{ \Theta( q-\Gamma_0^{(\ell)} - \xi_\ell - \K)}
{ E(q,\xi_\ell)h_\Delta(\xi_\ell)}
\prod_{j=1}^{R+1}
\frac{\Theta_\Delta(q-\gamma_j)}{\Theta_\Delta(q-\infty_j)} \times \le\{
\begin{array}{cl}
\ds \prod_{k=1}^n \frac {\Theta_\Delta(q-\infty^{(+)}_k)}{\Theta_\Delta(q-\infty^{(-)}_k)} & n\geq 0\\[18pt]
\ds \prod_{k=1}^{-n} \frac {\Theta_\Delta(q-\infty^{(-)}_k)}{\Theta_\Delta(q-\infty^{(+)}_k)} & n<0
\end{array}\ri.
\label{Tn}
\eea
\bea
&& \mathcal A = 2\le[ \oint_{a_1} \eta,\dots,\oint_{a_g} \eta\ri]^t\ \in \C^g \\
&&
\mathcal B =2 \sum_{j=1}^K t_j \mathfrak u(c_j) - 2 \le[ \oint_{b_1} \eta ,\dots, \oint_{b_g} \eta\ri]^t \ \in \C^g
\eea
Correspondingly the dual $(\nu, 1-\nu)$--differentials $\wavephi_n,\wavepsi_n$ are given by
\bea
\wavephi_n(q)&\& = \frac{T_n(q) \prod_{j=1}^K   \Theta_\Delta(q-c_j)^{-2i\pi t_j}}
{C_n\Theta_\Delta(q- \infty^{(-)}_{n+1})}
\Theta\le[
{-\mathcal A\atop -\mathcal B}
\ri]\le( q + \Gamma_n -  (2\nu-1)\K - \mathfrak X - \infty^{(-)}_{n+1} \ri) {\rm e}^{2i\pi \int_{p_0}^q \eta}h_\Delta(q)
\cr
\wavepsi_n(p) &\& = \frac{{T_n}^{-1}(p) \prod_{j=1}^K
  \Theta_\Delta(p-c_j)^{2i\pi t_j}}{\Theta_\Delta(p- \infty^{(+)}_{n+1})} 
\Theta\le[{\mathcal A\atop \mathcal B}
\ri]\le( p - \Gamma_n +   (2\nu-1)\K+ \mathfrak X - \infty^{(+)}_{n+1}
\ri) {\rm e}^{-2i\pi \int_{p_0}^p \eta}h_\Delta(p)\cr 
&\& \label{wavefunctions}
\eea
\be
C_n := \frac {\Theta\le[{\mathcal A\atop \mathcal B}
\ri]\le(  (2\nu-1)\K +\mathfrak X  - \Gamma_n\ri)\Theta\le[{\mathcal A\atop \mathcal B}
\ri]\le(     (2\nu-1)\K + \mathfrak X - \Gamma_{n+1} \ri)}
{ \Theta_\Delta\le(\infty^{(+)}_{n+1} - \infty^{(-)}_{n+1}\ri)}\ .
\ee
They are defined up to   rescaling by $\lambda, \lambda^{-1}$  respectively (corresponding to different choices of the common basepoint $p_0$ in the integrations above) and satisfy
\be
\res{\mathfrak \Infty^{(+)} }\wavephi_n \wavepsi_m = \delta_{mn} \label{Serreduality}
\ee
\et
{\bf Proof.}
The proof is a straightforward check that the proposed expression satisfies the defining properties in Eqs. (\ref{Kdiv1}).
The expression $T_n(q)$ in eq. \ref{Tn} has the following properties;
\begin{itemize}
\item it has a  tensor-weight of $\nu-\frac 1 2 $ in the variable $q$,
  i.e. can be written in a local coordinate as $f(z) (\d z)^{\nu-\frac
    1 2}$ and some multivaluedness around nontrivial cycles;
\item it has zeroes at $\Gamma_n$ and poles at $\mathfrak X$;
\item it is  {\bf projectively-independent} (i.e. independent up to a
  multiplicative constant) of  $\xi_j$ as long as they are not chosen
  so that the divisors appearing in the numerator are special. Indeed
  two different choices give functions with the same multivaluedness
  and the same divisor properties, hence proportional by a constant. 
\end{itemize}
The expression $G_n(p,q) = \frac{T_n(q)}{T_n(p)} \frac 1{E(p,q)}$ 
then has tensor weight $\nu$ in $q$ and $1-\nu$ in $p$ and the divisor properties
\be
(G_n(p,\bullet))\geq \Gamma_n - \mathfrak X - p\ ,\qquad
(G_n(\bullet,q))\geq -\Gamma_n + \mathfrak X - q
\ee
where $\Gamma_n$ are defined in eq. (\ref{divisToda}).
The remaining pieces of the formula make the final expression single valued and with the correct essential singularities. {\bf Q.E.D.}\par \vskip 5pt
As a corollary we derive the promised  relation between the spectral bidifferential and the Szeg\"o\ kernels
\bc
\label{sega}
The spectral bidifferential defined in Eq. \ref{specbidi} is given by 
\bea
S(p,q) &\&= \mathfrak K_n(p,q) \mathfrak K_n (q,p) =\cr
&\& = \frac {\Theta\le[{\mathcal A\atop \mathcal B} \ri]\le(p-q
  -\e_n\ri)\Theta\le[{\mathcal A\atop \mathcal B} \ri]\le(q-p -\e_n
  \ri)} 
{\Theta\le[{\mathcal A\atop \mathcal B} \ri]\le(-\e_n \ri)^2E^2(p,q)}\\
&\& \e_n :=\mathfrak X -  \Gamma_n -(2\nu-1)\K
\eea
\ec

In \cite{fay}, p. 26 we find that
\be
S(p,q)= \Omega(p,q) + \sum_{j,k=1}^g \frac {\pa^2 \ln \Theta}
{\pa_{\mathfrak u_j} \pa_{\mathfrak u_k}}(\e_n) \omega_j(p)\omega_k(q) 
\ee
where $\Omega$ is the {\bf normalized} fundamental bidifferential
(also known as Bergman kernel), such that $\oint_{a_j} \Omega \equiv
0$. 

Moreover, $S(p,q)$ is the square of the Szeg\"o\ kernel with complex
characteristics; specifically, if $\rho,\epsilon$ are the
(half)-characteristics of  $\e = \Gamma + (2\nu-1)\K - \mathfrak X +
\mathcal A + \tau \mathcal B$ 
\be
\e = 2 \rho + 2\tau \epsilon
\ee
then the above can also be rewritten as
\be
S(p,q) = \frac {\Theta\le[{\rho\atop \epsilon} \ri](p-q)
  \Theta\le[{\rho\atop \epsilon} \ri](q-p)}{\Theta\le[{\rho\atop
    \epsilon} \ri](0)^2E^2(p,q)} = \mathcal S_{\rho,\epsilon}(p,q)
\mathcal S_{\rho,\epsilon}(q,p)\ , 
\ee
where the Szeg\"o\ kernel with characteristics is defined by
\be
\mathcal S_{\rho,\epsilon}(p,q) = \frac {\Theta \le[ {\rho\atop
    \epsilon}\ri] (p-q)}{\Theta\le[{\rho\atop \epsilon}
  \ri](0)E(p,q)}\ . 
\ee

\section{Finite band recurrence relations and commuting (pseudo)  difference operators}
\label{sectFinite}

Let us  partition the polar divisor of $X$ into two {\bf disjoint}
subdivisors $\mathfrak X = \mathfrak X^{(+)} + \mathfrak X^{(-)}$ 
of degree $d$ and $R+1-d$ respectively :
\bea
(X)_- = -\sum_{j=0}^{d-1}\infty^{(+)}_j  - \sum_{j=0}^{R-d} \infty_j^{(-)} =:  -\mathfrak X\\
\deg (\mathfrak X) = R+1\ .
\eea
We will choose  the divisor $\mathfrak X^{(+)}$ as our {\bf dualization divisor}.

This means that according to the general scheme in Sect. \ref{gentoda}
we will choose all the points $\infty^{(+)}_j$ within $\mathfrak
X^{(+)}$ (hence the same symbol is used). Correspondingly, all the
points $\infty^{(-)}_j$ are chosen within $\mathfrak X^{(-)}$. 

In this expression the points $ \infty_j^{(\pm)}$ are {\bf not
  supposed} to be necessarily distinct (within the same subset), so
that we can consider 
poles of arbitrary order for $X$.

We will postulate that
\bea
&& \infty_{j+r}^{(-)} \equiv \infty_j^{(-)}\ ,\qquad r:= R-d+1\\
&& \infty_{j+d}^{(+)} \equiv \infty_j^{(+)}\
\eea
and assume that $X$ has at least two distinct poles ($r\geq 1 $); the
modifications for the case of a single pole are left to the reader. 
We also fix a third kind differential $\eta$ as in Sect. \ref{preflat}.

Define the divisors
\bea
\mathfrak U_n &\& :=\le\{
\begin{array}{cc}
\ds  -\Gamma +\mathfrak X -\sum_{j=1}^{n+1}\infty^{(+)}_j +  \sum_{\ell=1}^{n} \infty^{(-)}_\ell & n \geq 0 \\
 \ds - \Gamma +\mathfrak X +\sum_{j=n+1}^{0}\infty^{(+)}_j  - \sum_{\ell = n}^{0} \infty^{(-)}_\ell & n < 0
 \end{array}
 \ri.\\
&\& \deg(\mathfrak U_n) =- (2\nu-1)(g-1) - 1\\
\mathfrak V_n &\& :=\le\{
\begin{array}{cc}
 \ds \Gamma -\mathfrak X +  \sum_{j=1}^{n}\infty^{(+)}_j - \sum_{\ell=1}^{n+1} \infty^{(-)}_\ell & n\geq 0 \cr
 \ds \Gamma -\mathfrak X - \sum_{j=-n}^0 \infty^{(+)}_j +  \sum_{\ell=-n+1}^0 \infty^{(-)}_\ell & n<  0
 \end{array}\ri .\\
&\& \deg(\mathfrak V_n) = (2\nu-1)(g-1)-1
\eea
In the formul\ae\ above it is understood that the sums are zero if the ranges are empty.
The main point of these definitions is that
\be
\mathfrak U_n = -\Gamma_n +\mathfrak X - \infty_{n+1}^{(+)}\ ,\qquad
\mathfrak V_n = \Gamma_n -\mathfrak X - \infty_{n+1}^{(-)}\ .
\ee
and these, in view of Sect. \ref{gentoda},  are precisely the divisors
characterizing (up to scalar multiplication)\footnote{The essential
  singularities described in Sect. \ref{preflat} are implied.} 
  $\wavepsi_n$ and $\wavephi_n$
\be
\begin{array}{cl}
(\wavepsi_n) \geq \mathfrak U_n & \hbox { a $(1-\nu)$--differential}\\
(\wavephi_n)\geq \mathfrak V_n & \hbox{ a $\nu$--differential}
\end{array}
\ee

These two sequences span vector spaces dual to each other under {\bf Serre's} duality;
\be
\res{\mathfrak X^{(+)}} \wavepsi_n \wavephi_m =  \delta_{mn}\ \label{Serre}
\ee
Another crucial point that motivates the choice of  twisting divisors is that now
\bea
&& \C\{\wavepsi_{j}\}_{d-R\leq j\leq d} = \mathcal H_{1-\nu} (-\Gamma_0)\\
&& \C\{\wavephi_{j}\}_{-d\leq j\leq R-d} = \mathcal H_{\nu} (\Gamma_0-2\mathfrak X)
\eea
and more generally
\bea
&& \C\{\wavepsi_{j}\}_{d-R+n\leq j\leq n+ d} = \mathcal H_{1-\nu} (-\Gamma_n)\\
&& \C\{\wavephi_{j}\}_{n-d\leq j\leq n+ R-d} = \mathcal H_{\nu} (\Gamma_n-2\mathfrak X)
\eea
and hence we can conveniently choose them as components of the vectors
$\bpsi_n, \bvarphi_n$  used in the general construction. 

 In this fashion, the vectors $\bpsi_n$ and $\bvarphi_n$ are {\bf
   windows} of consecutive $R+1$ elements within a pair of infinite
 vectors 
 
\parbox{9cm}{
$$
\Psi := \le[\begin{array}{c}
\vdots\\
\wavepsi_{n-R+d}\\
\vdots\\
 \vdots\\
 \wavepsi_n\\
 \vdots\\
  \wavepsi_{n+d}\\
   \vdots\\
 \vdots
\end{array}\ri ]
\begin{picture}(50,50)(0,0)
\put(0,0){\line(1,0){40}}
\put(0,50){\line(4,-1){40}}
\put(0,-40){\line(4,-1){40}}
\end{picture}
\le[
\begin{array}{c}
\vdots\\
\vdots\\
\wavephi_{n-d}\\
\vdots\\
\wavephi_n\\
\vdots\\
 \vdots\\
\wavephi_{n+R-d}\\
 \vdots\end{array}\ri]=:\Phi 
 $$}
\parbox{6cm}{ \bd
 The vectors $\Psi,\Phi$ will be called the {\bf wave--vectors}. The  dual {\bf windows} are defined by
 \bea
 \bpsi_n := \le[\wavepsi_{n-R+d},\dots, \wavepsi_{n+d}\ri]^t\ ,\cr
 \bvarphi_n:=\le[\wavephi_{n-d},\dots, \wavephi_{n+R-d}\ri]^t\ ,\nonumber
 \eea
 where the entries, depending on the base divisor $\Gamma = \Gamma_0$
 and the line bundle associated to the differential $\eta$ are given
 by the expressions in Thm. \ref{thm_nu_Cauchy}. 
 \ed}
 
\parbox{8cm}{
\bp
 The wave vectors satisfy a finite band recurrence relation
 \be
 X \Psi = \X \Psi\ ,\qquad X\Phi^t = \Phi^t \X
 \ee
 where the doubly infinite matrix $\X$  has a finite band structure of the form indicated on the right, and  $\a_j(n):= \res{\mathfrak X^{(+)}} \wavephi_{n+j} X \wavepsi_n$.
\ep} \parbox{6cm}{\resizebox{5cm}{!}
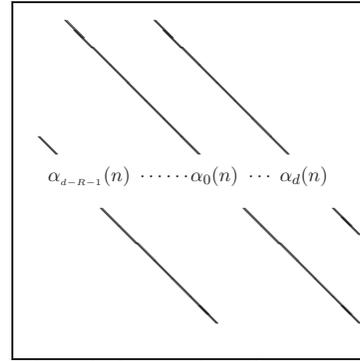
{
 $$
 \begin{picture}(200,200)(-150,-100)
 \put(-150,0){$\X=$}
\put(-100,-100){\framebox(200,200)}
\put(0,0){$\a_0(n)\ \cdots$}
\put(-80,0){$\a_{_{d-R-1}}(n)\ \cdots\cdots$}
\put(50,0){$\a_{d}(n)$}
\put (5,15){\line(-1,1){75}}
\put (5.5,15){\line(-1,1){75}}
\put (4.5,15){\line(-1,1){75}}
\put (-75.5,15){\line(-1,1){10}}
\put (-75,15){\line(-1,1){10}}
\put (-74.5,15){\line(-1,1){10}}
\put (55.5,15){\line(-1,1){75}}
\put (55,15){\line(-1,1){75}}
\put (54.5,15){\line(-1,1){75}}
\put(30.5,-15){\line(1,-1){65}}
\put(30,-15){\line(1,-1){65}}
\put(29.5,-15){\line(1,-1){65}}
\put(80.5,-15){\line(1,-1){15}}
\put(80,-15){\line(1,-1){15}}
\put(79.5,-15){\line(1,-1){15}}
\put(-50.5,-15){\line(1,-1){65}}
\put(-50,-15){\line(1,-1){65}}
\put(-49.5,-15){\line(1,-1){65}}
 \end{picture}
 $$}}

 {\bf Proof.}
 The fact that $\Psi, \Phi$ solve transposed recurrence relations is immediate from the residue--pairing
 \be
\res{\mathfrak X^{(+)}}\Psi \Phi^t  = \1\ ,\qquad
\res{\mathfrak X^{(+)}} X\Psi \Phi^t = \X
 \ee
 where $\1$ is the infinite identity matrix.
 The shape of the matrix $X$ follows from inspection of the divisor properties of $\wavepsi_n$ and $X \wavepsi_n$.
 {\bf Q.E.D.}\par \vskip 5pt
Keeping this in mind we can prove
\bp
\label{commCDI}
The Christoffel--Darboux pairing $\Amat_n$ is given by the  non-zero $(R+1)\times (R+1)$ block in
\be
\Amat_n := [\Pi_n,\X]\ ,
\ee
where $\Pi_n = {\rm diag}(\dots, \dots, 1,0,\dots)$ is the projector
up to $n$ (i.e. the zero entries on the diagonal start at the entry
$(n+1,n+1)$). 
\ep
{\bf Proof.}
Using the definition of the matrix $\Amat_n$
\be
(X(p)-X(q)) \mathfrak K_n(p,q) =\bvarphi_n^t(q) \Amat_n \bpsi_n(p)
\ee
we find that the entries of $\Amat_n$ are given by
\bea
(\Amat_n)_{a,b}= \res{p\in \mathfrak X^{(+)}} &\& \res{q\in \mathfrak
  X^{(+)}} (X(p)-X(q)) \mathfrak K_n(p,q) \wavephi_{a}(p) \wavepsi_{b}
(q)\cr 
&& \qquad a=n-d,\dots,n+R-d \ ; \ \ b= n-R+d,\dots, n+d
\eea

Note that the order of the residues is irrelevant because the
integrand is regular on the diagonal $p=q$. Using relations
(\ref{projphipsi}) we conclude that the nonzero entries are 
\be
\Amat_{ab} = \le\{
\begin{array}{cc}
\ds \res{\mathfrak X^{(+)}} X\wavephi_a \wavepsi_b &  b\geq n+1\ ,\  a\leq n-1\\
\ds -\res{\mathfrak X^{(+)}} X\wavephi_a\wavepsi_b & b\leq n-1\ ,\ a \geq n+1\
\end{array}
\ri.
\ee
Explicitly, using the notation $\X_{nm} = \res{\mathfrak X^{(+)} } \varphi_{n+k} X \psi_n = \a_k(n)$, we have
(recall also that $\a_k(n) \equiv 0$ for $k>d$ and $k<d-R$)
\bea
\Amat=\Amat_{n} =  -\le[
\begin{array}{cccc|cccc}
&&&&  \a_d(n-d)&&\\
&&&&\vdots  &\hspace{-10pt}\ddots &\\
&&&  &\cdots &\cdots&\a_d(n)\\
\hline
-\a_{d\!-\!R\!-\!1}(n+1) & \cdots &&-\a_{-1}(n+1)  &&&\\
& \ddots & \ddots & \vdots &  & &&\\
&& \ddots & \vdots &  & &&\\
&&&\hspace{-70pt}\ddots &&&&\\
&&&-\a_{d\!-\!R\!-\!1}(n\!+\!d\!-\!R) & &&&
\end{array}
\ri]
\eea
This concludes the proof. {\bf Q.E.D.} \par \vskip 5pt

Since all basis elements $\bpsi_n,\bvarphi_n$ and the coefficients of
$\Amat_n$ have been expressed in the most direct way as Theta
functions or residues thereof, we have achieved our goal of 
providing a completely explicit expression for the solution of the
inverse spectral problem described by formula (\ref{Laxmatrix})  and
Section \ref{sectLax}. 

Of course one should consider only one member of the sequence, without reference to the full sequence: so, for example, we can identify the matrix in (\ref{Laxmatrix}) with the zeroth term.

\subsection{Lax and ladder matrices}
As it was explained in
Sect. \ref{gentoda}, we have a {\bf sequence of Lax matrices}
$\{A_n(x)\}_{n\in\Z}$ and an intertwining sequence of {\bf ladder
  matrices} $\{C_n(x)\}_{n\in\Z}$ such that 
\be
A_{n+1} (x) = C_n(x)^{-1} A_n(x) C_n(x)\ .
\ee
The ladder matrices $C_n$ are linear in $x$ as follows from the
explicit formula in Sect. \ref{sectspec}. In this context the formula
reads 
\bea
C_n(x) = \res{\mathfrak X^{(+)}} \frac{\bpsi_{n+1}(p)\bvarphi_n(p)^t \mathbb K_n}{X(p)-x}\\
\bpsi_{n+1}(p) = C_n(X(p)) \bpsi_n(p)\ .
\eea

The reader should check that they have companion--like form as in \cite{BEH}, where the coefficients of $\X$ appear in the nontrivial row (column) of the ladder matrix. 

The situation is very similar to the recurrence relation satisfied by
orthogonal polynomials and generalization thereof \cite{BEH}: in that
case the recurrence relations are typically of Hessenberg form ($d=1$
in our setting). 

However in the case of (generalized)  orthogonal polynomials,  the
ladder matrices induce  Schlesinger transformations for the associated
Riemann--Hilbert problem, whereas here they have simply the meaning of
an elementary twisting of the divisor. 
Note that the matrix representing multiplication by $X$ in the
infinite basis of the wave--functions $\Psi$ (or, dually, $\Phi$) is a
finite--band matrix by construction; on the other hand multiplication
by $Y$ does not result in a finite band matrix, namely the matrix 
\be
\Y_{nm} := \res{\mathfrak X^{(+)}} Y \wavepsi_m\wavephi_n
\ee
is in general a ``full'' doubly-infinite matrix; it has a finite
number of nonzero supradiagonals (corresponding to the degree of the
subdivisor of $\mathfrak Y$  supported at $\mathfrak X^{(+)}$) but in
general the part below the diagonal is not finite-band. 

Nonetheless the following matricial identity holds
\be
[\X,\Y]=0
\ee
which makes sense entry-wise
since, $X$ being finite band,  the commutator involves only a finite
number of terms. The commutativity clearly follows from the fact that
the two matrices represent commuting multiplication operators.Thus we
are looking at a pair of {\bf commuting pseudo-difference operators},
where $\X$ is a bona-fide difference operator, while $\Y$ is not. 
The only case in which $\Y$ is of finite band structure as well is if
the divisor of poles of $Y$ coincides with the polar divisor of $X$
(although of different multiplicity in general). 
We remark however that in the case of pseudo-difference operators, even if the matrix $\Y$ has no obvious shape it has  nonetheless  a hidden {\bf rank condition}. 

Indeed, under some additional genericity assumptions we can factor
$\Y$ into the inverse of a lower-triangular matrix and a finite--band
matrix. 

To see this, let us separate the poles of  $Y$ into the poles
$\mathfrak Y_x$ that are also poles of $X$ and the ``other''  poles
$\mathfrak Y_o$ of  degrees $d_x$ and $d_o$ respectively. 

In a generic situation, we can find a linear combination
$\wt{\wavepsi}_n\in \C\{\wavepsi_n, \dots, \wavepsi_{n-d_o}\}$ whose
divisor exceeds $\mathfrak Y_o$ and hence  $ Y\wt{\wavepsi}_n$ will be
a linear combination of $\wavepsi_{n+j}$, for $|j|<d_x$; the actual
shape depends on how the poles $\mathfrak Y_x$ are distributed into
the divisors $\mathfrak X^{(\pm)}$. 

Setting $\mathfrak Y_x^{(\pm)} = \mathfrak Y_x\cap \mathfrak
X^{(\pm)}$ and letting $m_\pm$ be the respective degrees, we  see that
there is a lower--triangular matrix $L$ with $d_o$ subdiagonals and a
finite--band matrix $H$ with $m_+$ supradiagonals and $m_-$
subdiagonals such that 
\be
YL\Psi = H \Psi\ \ \Rightarrow \ \ \Y = L^{-1} H\ .
\ee
This implies that all the submatrices below the main diagonal have
rank at most $d_o$ and this is our ``hidden'' rank condition. 
\br
A more general construction is needed in application to large degree
asymptotics of certain biorthogonal polynomials (we will pursue this
in a different publication  \cite{BertoMishaJacek}). In this case the
twisting of the base divisor is performed in a different way so that
{\bf both} $\X$ and $\Y$ are pseudo difference operators subject to
similar rank conditions. 
\er
\subsection{Riemann--Hilbert problems}
Although it is  outside of the scope of this paper, we would like to explain what makes this investigation
of potential relevance in a study of large degree asymptotics of (multi)--orthogonal polynomials.

We start with the observation that the windows $\bpsi_n,\bvarphi_n^t$
are eigenvectors of a matrix $A_n(x)$ which depends only on the value
$x=X(p)$; since there are $R+1=\deg (X)$ other points $p_1(x),\dots,
p_{R+1}(x)\in \L$ (generically distinct) with the same
$X$--projection, the evaluation at those points provides a basis of
eigenvectors. The points (and so the eigenvectors) are distinct away
from the ramification divisor of the map $X:\L\to \C$ and hence the
sections $p_i:\C\to\L$ are well defined only on a suitable simply
connected domain obtained by removing some cuts originating at the
branch-points of the $X$--projection. 

The matrices
\bea
P_R(x)&\& := \le[\frac{\bpsi_n(p_1(x))}{\d X(p_1(x))^{1-\nu} }{\rm
  e}^{-2i\pi \int^{p_1(x)}\eta},\dots, \frac {\bpsi_n(p_{R+1}(x))}{\d
  X(p_{R+1}(x))^{1-\nu}} {\rm e}^{-2i\pi \int^{p_1(x)}\eta}\ri]\\ 
P_L(x)&\& := \le[
\begin{array}{c}
\frac{\bvarphi_n^t(p_1(x))\mathbb K_n}{\d X(p_1(x))^\nu} {\rm
  e}^{2i\pi \int^{p_1(x)}\eta}\\ \vdots\\
\frac{\bvarphi_n^t(p_{R+1}(x))\mathbb K_n} {\d X(p_{R+1}(x))^\nu} {\rm
  e}^{2i\pi \int^{p_1(x)}\eta} 
\end{array}\ri]
\eea
are inverses of each other.
They solve a Riemann--Hilbert problem with quasipermutation
monodromies around the branch-points of $X$ (due to the permutation of
columns and to the multivaluedness of 
columns as functions on the spectral curve itself) and diagonal
multivaluedness around the $X$--projection of the poles of $\eta$ 
(due to logarithmic singularities
). It would not be difficult, but too long, to spell out in detail the
Riemann-Hilbert data, as they include some growth conditions at
infinity and at the branch-points. Note in general that we can expect
singularities at a branch-point $x_o$ of order $k$ of the form 
$(x-x_o)^{\frac {\nu-1}{k}}$ for $P_R$ and of the form
$(x-x_o)^{-\frac {\nu}k}$ for $P_L$, or combination of singularities
of this type if there are more than one ramification points on the
spectral curve above the same branch-point. 

Riemann--Hilbert problems of this sort have been used in
\cite{BertoMoAsympt,BertoMishaJacek} in the asymptotic analysis of certain
(bi)orthogonal polynomials 
(for the case $\nu=\frac 1 2 $).

 The point of contact between the above RHP and the ones satisfied by
 (multi)orthogonal polynomials in the asymptotic regime is that such
 problems with 
 quasi--permutation monodromies appear when the original RHP is
 ``normalized'' by the use of a suitable collection of $g$--functions
 \cite{DKMVZ}. 
\section{Commuting (pseudo)--difference operators in duality related to the two--matrix model}
\label{sectComm}
We consider now a  particular case which is of relevance for the
asymptotic analysis of the biorthogonal polynomials for the so--called
``two--matrix model'' \cite{EynMehta, BEH}; 
we will remark later on what are the choices of the tensor weights and
divisors which are more strictly relevant to that situation. 

The restriction will be that $X$ and $Y$ share the same polar divisor in the specific form
\cite{Berto2Toda}
\be
(X)\geq -\infty_x - d_2\infty_y \ ,\qquad
(Y)\geq -d_1\infty_x - \infty_y\ .
\ee
We use $\Infty^{(+)} = \infty_x$, $\Infty^{(-)} = \infty_y$ and the
same general framework used earlier, with a (generic) divisor $\Gamma$
of degree $(2\nu-1)(g-1) + d_2$ and the third--kind differential
$\eta$. 

 The wave-vectors are then characterized (up to constants) by the
 formul\ae\  in Thm. \ref{thm_nu_Cauchy} suitably specialized; for
 reader's convenience we recall 
 the divisor properties
\bea
(\wavepsi_n) \geq -\Gamma -n\infty_x + (n+d_2)\infty_y\ ,\ \ \
(\wavephi_n)\geq \Gamma +(n-1)\infty_x - (n+1+d_2)\infty_y\\
\res{\infty_x} \wavepsi_n\wavephi_m = \delta_{mn}\ .
\eea
Since the functions $X,Y$ share the polar divisor, the sequence of
wave-functions also satisfies a finite band $Y$--recurrence relation
and hence the matrices 
$\X,\Y$ can be thought of as  {\bf two commuting difference operators}\cite{KrichNov};
denoting
\bea
\Y_{m,n} :=  -\res{\infty_x} Y \wavephi_{m} \wavepsi_n\\
Y \Psi = -\Y \Psi\ ,\qquad
Y\Phi^t = -\Phi^t \Y
\eea
we see that the two matrices $\X,\Y$ are finite--band Hessenberg
matrices with $d_2$ subdiagonals for $\X$ and $d_1$ supradiagonals for
$\Y$. 

On the other hand we could have
switched the r\^oles of $X$ and $Y$, used a (generic) divisor $\wt
\Gamma$ of degree $(2\wt \nu-1)(g-1)+d_1$ and a differential $\wt\eta$
and repeated 
 the whole construction so as to get another pair of sequences of wave--functions
\bea
(\wt{\wavepsi}_n)&\& \geq -\wt\Gamma -n\infty_y + (n+d_1)\infty_x\ ,\ \ \
(\wt{\wavephi}_n)\geq \wt\Gamma +(n-1)\infty_y - (n+1+d_1)\infty_x\\
&&\res{\infty_y} \wt{\wavepsi}_n\wt{\wavephi}_m = \delta_{mn}\\
&& \wt \Y_{nm} = \res{\infty_y} Y \wt{\wavepsi}_n\wt{\wavephi}_m \ ,\qquad
\wt \X_{nm} = -\res{\infty_y} X \wt{\wavepsi}_n\wt{\wavephi}_m
\eea
It is clear that in general the matrices $\X,\Y$ and their tilde-counterparts have nothing in common except the shape. In
the formal asymptotics 
 of biorthogonal polynomials the two matrices should however be the same \cite{BEH}: we will see below that this implies certain constraints on the divisors $\Gamma, \wt\Gamma$ and the differentials  $\eta,\wt \eta$.

Suppose that
\be
\eta + \wt \eta = d F = \hbox {exact differential}\ ,\qquad
\Gamma + \wt \Gamma - \mathfrak X  - \mathfrak Y \equiv (\nu+\wt\nu-1) \mathcal C\ \label{nucano}
\ee
where $\mathcal C$ is a canonical divisor. This means that there
exists a $(\nu+\wt \nu-1)$--differential whose divisor is the one
above. 
Also we assume that $\eta , \wt\eta $ have the same polar divisor (in particular opposite residues).
Under these conditions we have
\bp
\label{laplace}
If the divisors $\Gamma, \wt \Gamma$ and differentials $\eta, \wt \eta$ are {\bf dual} in the sense of
eq. (\ref{nucano})  then there exists a $(\nu,\wt \nu)$
--bidifferential $\mathfrak L$ which we call the {\bf Laplace kernel}
and a $(1-\nu,1-\wt \nu)$--bidifferential $\wh{\mathfrak L}$ which we
call the {\bf co-Laplace kernel} with the properties 
\bea
&& (\mathfrak L(p,q))_p\geq \Gamma -\infty_x -d_2\infty_y -q\ ,\qquad
(\mathfrak L(p,q))_q \geq \wt\Gamma -d_1\infty_x -\infty_y -p \\
&& (\wh{\mathfrak L}(p,q))_q\geq -\Gamma +\infty_x + d_2\infty_y -p\ ,\ \
(\wh{\mathfrak L} (p,q))_p \geq -\wt\Gamma + d_1\infty_x + \infty_y -q
\eea
Along the diagonal $p=q$ they behave as
\bea
\mathfrak L \sim \d z^\nu \d{ z'}^{\wt\nu} \frac {f(z)}{z-z'} + \dots\\
\wh {\mathfrak L} \sim  \d z^{1-\nu} \d{ z'}^{1-\wt\nu} \frac {\wh f(z)}{z-z'} + \dots
\eea
where $\omega := f(z) \d z^{\nu+\wt\nu-1}$ and $\wh \omega := \wh f(z)
\d z^{1-\nu-\wt \nu}$ are invariantly defined differentials of the
indicated weights. 
\ep
{\bf Proof}
Similarly to the construction of the $\nu$--Cauchy kernel, we split the divisors $\Gamma, \wt\Gamma$ into
\bea
\Gamma = \sum_{j=1}^{d_2+1} \gamma_j +\sum_{a=1}^{2\nu-1} \overbrace{\Gamma^{(a)}}^{\deg = g-1}\\
\wt \Gamma = \sum_{j=1}^{d_1+1} \wt \gamma_j +\sum_{a=1}^{2\wt \nu-1} \overbrace{\wt \Gamma^{(a)}}^{\deg = g-1}
\eea
and choose $2\nu-1$ points $\xi_a$ and $2\wt \nu-1$ points $\wt \xi_b$
(again, the formul\ae\ will depend only projectively on those). 
We rewrite formul\ae\ (\ref{Feta}, \ref{Tn}, \ref{wavefunctions}) specializing them twice: 
\begin{itemize}
\item
the first time with $\infty^{(+)}_n \equiv \infty_x,\ \infty^{(-)}_n \equiv \infty_y, \mathfrak X = \infty_x + d_2 \infty_y$;
\item the second time with  $\wt \infty^{(+)}_n \equiv \infty_y,\ \wt \infty^{(-)}_n \equiv \infty_x, \mathfrak Y = \infty_y + d_1 \infty_x$.
\end{itemize}
%


 Duality (\ref{nucano}) implies that 
\bea
\wt{\mathcal A} =-\mathcal A\ ,\ \ \wt{\mathcal B} = -\mathcal B\ ,\ \wt t_j = -t_j\ ,\ 
\mathfrak u\le( \Gamma + (2\nu-1)\K - \mathfrak X\ri)  =  \mathfrak u\le(-\wt \Gamma - (2\wt \nu -1) \K +\mathfrak Y
\ri)
\eea
so that 
\be
F_{\eta,0}(p,q) =  F_{\wt\eta,0} (q,p)
\ee
with the quantities being defined as in eq. (\ref{Feta}) and the above specializations.
Then a direct inspection shows that ($T_0$ defined as part of (\ref{Tn}))
\bea
\mathfrak L(p,q) = \frac {T_0(p)\wt T_0(q)}{E(p,q)} F_{\wt \eta,0}(p,q){\exp }\bigg({\overbrace{-2i\pi \int_p^q \eta + 2i\pi F(q)}^{2i\pi F(p) -2i\pi  \int_q^p \wt \eta}}\bigg)\\
\wh{\mathfrak L}(p,q) = \frac{ F_{\eta,0}(p,q)}{T_0(p)\wt T_0(q) E(p,q)}{\rm e}^{2i\pi \int_p^q \eta - 2i\pi F(q)}
\eea
The $(\nu+\wt\nu-1)$--differential and $(1-\nu-\wt\nu)$--differential  advocated for in the proposition are (up to multiplicative constant)
\be
\omega(p):= T_0(p)\wt T_0(p) {\rm e}^{2i\pi F(p)}  \ ,\qquad \wh \omega (p) = \frac 1{\omega(p)}\ ,\label{omegas}
\ee
where the fact that these expressions are single--valued follows once more from (\ref{nucano}).
{\bf Q.E.D.}\par\vskip 5pt
\bc
The normalizations of the four sequences of wave--functions can be chosen in such a way that
\bea
&& \res{\xi= \infty_{x,y}} \mathfrak L(\xi,q) \wavepsi_n(\xi) = \omega(q)\wavepsi_n(q) = \wt{\wavephi}_n(q)\\
&& \res{\xi= \infty_{x,y}} \wh{\mathfrak L}(\xi,p) \wavephi_n(\xi) =
\wh \omega(p) \wavephi_n(p) = \wt{\wavepsi}_n(p)\\ 
&& \res{\xi= \infty_{x,y}} \mathfrak L(p,\xi) \wt{\wavepsi}_n(\xi) =\omega(p)  \wt \wavepsi_n(p)=  \wavephi_n(p)\\
&& \res{\xi= \infty_{x,y}} \wh{\mathfrak L}(q,\xi)
\wt{\wavephi}_n(\xi) =\wh \omega(p) \wt\wavephi_n(q) =  \wavepsi_n(q) 
\eea
\ec
{\bf Proof.}
We first note that the differentials appearing in all the  residues
above are {\bf meromorphic} since the essential singularities (by
construction) cancel out. The only poles are {\em a priori} at
$\infty_{x,y}$ and hence the sum over the two residues is (minus) the
residue along the diagonal. Thus the statement of the theorem, keeping
into account that $\omega \wh \omega \equiv 1$,  amounts to checking
that 
\be
\wavepsi_n(p) = \wh \omega(p) \wt \wavephi_n(p)\ ,\qquad
\wavephi_n(p) = \omega(p) \wt \wavepsi_n(p)\label{dddd}\ .
\ee
The check is a straightforward computation using the explicit
expressions (\ref{omegas}) and the expressions for the two dual
sequences derived from the suitable specializations of
eqs. (\ref{wavefunctions}): the tilded sequences defined in (\ref{dddd}) coincide with those obtained by specialization of (\ref{wavefunctions}) up to a rescaling $\wt \rho_n \to \lambda_n \wt \rho_n,\ \wt \pi_n\to \frac 1 {\lambda_n} \wt \pi_n$ (which leaves invariant the duality (\ref{Serreduality})).
{\bf Q.E.D.}\par \vskip 5pt
\bc
If the duality (\ref{nucano}) is satisfied then the matrices
representing multiplication by $X$ and $Y$ are the same (up to a
transposition and a sign) in the two dual bases. 
\ec
{\bf Proof.}
Indeed
\bea
& &\X_{nm} = \res{\infty_x} X\wavepsi_n \wavephi_m =  \res{\infty_x}
X\omega \wavepsi_n \wh \omega \wavephi_m =  \res{\infty_x} X \wt
\wavephi_n \wt \wavepsi_m = -  \res{\infty_y} X \wt \wavephi_n \wt
\wavepsi_m = -\wt \X_{mn}\\ 
&& \Y_{nm} = \res{\infty_x} Y\wavepsi_n \wavephi_m =  \res{\infty_x}
Y\omega \wavepsi_n \wh \omega \wavephi_m =  \res{\infty_x} Y \wt
\wavephi_n \wt \wavepsi_m = -  \res{\infty_y} Y \wt \wavephi_n \wt
\wavepsi_m = -\wt\Y_{mn} 
\eea
{\bf Q.E.D.}\par \vskip 5pt

There are two Lax-matrices (see \cite{BEH} for the analysis for
biorthogonal polynomials): $A_n(x)$ of size $(d_2+1)\times (d_2+1)$
and $B_n(y)$ of size 
$(d_1+1)\times (d_1+1)$ which share the characteristic polynomial (by construction).

In addition, due to Corollary  \ref{sega} and the duality \ref{nucano}
the two spectral bidifferential coincide. 

This means that if we denote by $A(x)$ the Lax matrix reconstructed using $X(p)$ as spectral parameter (of dimension $d_2+1$) and $B(y)$ the $(d_1+1)^2$ Lax matrix obtained by using instead $Y(p)$ as spectral parameter and $X(p)$ as eigenvalue, we have  the identity
\bea
&& \det(x\1_{d_1+1}- B(y)) = c \det (y\1 _{d_2+1} -A(x))\\
&& \frac {\d x\d x'}{(x-x')^2} \frac{\tr \le((\wt{A - y})(x)(\wt{A- y})(x')\ri)}{\tr \le((\wt{A - y})(x)\ri)\tr \le((\wt{A - y})(x')\ri)} = \cr
&&\ \ \ = \frac {\d y\d y '}{(y-y')^2} \frac{\tr \le((\wt{B - x})(y)(\wt{B- x })(y')\ri)}{\tr \le((\wt{B - x})(y)\ri)\tr \le((\wt{B - x})(y')\ri)} 
\eea
\paragraph{Conclusion.}
We conclude the section by pointing out the specialization that will
be of use in the study of the asymptotics for biorthogonal polynomials 
appearing in the two--matrix model; indeed, the choice of the differentials $\eta, \wt \eta$ above was too generic.

The relevant case would be $\nu = \wt \nu = \frac 1 2 $; in this case
$\Gamma, \wt \Gamma$ have degrees $d_2,d_1$ respectively and one 
should choose them as $\Gamma = d_2\infty_y,\ \wt \Gamma =
d_1\infty_x$. The differential $\eta$ is then defined by $2i\pi \eta =
N Y\d X$ and the  dual one by 
$2i\pi \wt\eta   = N X\d Y$ so that in (\ref{nucano}), $2i\pi F = NXY$.

The parameter $N$ in the biorthogonal polynomial context is  a large parameter (corresponding to the size of the underlying matrix model) and the degree of the polynomials whose asymptotics we are interested in, is $n = N+r$ with $r\in \Z$ arbitrary but not scaling with $N$, i.e. bounded.

In the spirit of this paper we should think of $N$ as a deformation of the {\em line bundle} $\eta$ while the ``fluctuations around the Fermi level'' (using a common analogy in the physically oriented literature) should be identified with $r$ and give rise to the generalized Toda lattice (in this case it is the $2$--Toda lattice).

\end{document}